\documentclass[12pt,preprint2]{aastex}

\shorttitle{On the solar signal at 220.7}

\shortauthors{A. Jim\'enez \& R.A. Garc\'\i a}

\begin{document}

\title{On the solar origin of the signal at 220.7$\mu$Hz: A possible component of a g mode?}

\author{A. Jim\'enez\altaffilmark{1}, R. A. Garc\'\i a\altaffilmark{2}}

\altaffiltext{1}{Instituto de Astrof\'\i sica de
Canarias, E-38205 La Laguna, Tenerife, Spain}
\altaffiltext{2}{Service d'Astrophysique CEA/DSM/DAPNIA, CE Saclay, 
91191 Gif-sur-Yvette CEDEX, France}

\begin{abstract}
Gravity modes in the Sun have been the object of a long and difficult search in recent decades. Thanks to the data 
accumulated with the last generation of instruments (BiSON, GONG and three helioseismic instruments aboard SoHO), scientists 
have been able to find  signatures of their presence. However, the individual detection of such modes remains evasive. 
In this article, we study the signal at 220.7 $\mu$Hz which is a peak that is present in most of the helioseismic data 
of the last 10 years. This signal has already been identified as being one component of a g-mode candidate detected 
in the GOLF Doppler velocity signal. The nature of this peak is studied in particular using the VIRGO/SPM instrument 
aboard SoHO. First we analyse all the available instrumental data of VIRGO and SoHO (housekeeping) to reject any possible 
instrumental origin. No relation was found, implying that the signal has a solar origin. Using Monte Carlo simulations, 
we find, with more than 99$\%$ confidence level, that the signal found in VIRGO/SPM is very unlikely to be due to pure noise.
\end{abstract}

\section{Introduction}
Helioseismology has probed the interior of the Sun over the last three decades. Combining the information provided by 
several hundred pressure-driven modes (p modes), it has been possible to put  constraints on our knowledge of the 
structure and the dynamics of the solar interior \citep{JCD2002,ThoJCD2003}. Unfortunately, only a small fraction of 
these modes reaches the solar core. Moreover, due to the increase in  sound-speed velocity with depth, these modes 
give little information on the deeper layers as they spend less time there than in the convection zone. Let us take as an 
example the internal rotation rate of the Sun: the rotation profile is very well known in the convective zone 
\citep{ThoTOO1996,1998ApJ...505..390S,HowJCD2000,2000ApJ...541..442A}, while the uncertainties grow in the radiative 
zone and towards the core of the Sun \citep{1994ApJ...435..874J,ElsHow1995,CouGar2003,ChaSek2004,GarCor2004}. In order 
to put new constraints inside the solar core other kinds of modes are needed: the gravity (g) modes. For example, by 
measuring just a few of such modes, information on the core rotation rate can undoubtably be obtained 
\citep{2008A&A...484..517M}, whereas   the deepest layers that could be probed using only p-modes are around 
0.2 $R_{\odot}$ \citep{2008SoPh..tmp...43G}.
 
Gravity (g) modes are buoyancy-driven modes that have the advantage of propagating inside the entire radiative 
region. However, these waves become evanescent in the convective zone and reach the solar surface with tiny 
amplitudes preventing us to detect them easily (see for example \citet{Belkacem08} and references therein). Indeed, 
several claims for g-mode detections have been made in the past \citep{1983Natur.306..651D,1988IAUS..123...79P,1995Natur.376..139T}; 
however, modern and better data sets cannot confirm them. 

In 1995 was launched the Solar and Heliospheric Observatory (SoHO),  one of whose scientific objectives 
was the detection and characterization of gravity modes \citep{DomFle1995}. Recently, using data from the Global 
Oscillation and Low Frequency (GOLF) instrument \citep{GabGre1995}, the signature of the asymptotic properties of 
  $\ell$=1 dipole g modes has been measured with a high confidence level \citep{2007Sci...316.1591G,2008AN....329..476G}. 
This signal was also found \citep{2006ESASP.624E..23G} using photometric data from the Variability of solar 
IRradiance and Gravity Oscillations (VIRGO) experiment \citep{1995SoPh..162..101F}. Even if certain constraints 
can be imposed on the structure \citep{2008SoPh..tmp...55G} and dynamics \citep{2007Sci...316.1591G} of the 
solar core thanks to the study of these asymptotic properties, it is extremely important to detect individual 
g modes.  After 10 years of observations, the level of noise at 200 $\mu$Hz has been established at 
$\sim$ 4.5 mm/s when individual peaks are looked for and at 1.5 mm/s when this research is done for multiplets 
\citep{2006soho...18E..22E}. Indeed some peaks and patterns could be identified as potential gravity modes or
 mixed modes above  the  noise level \citep{GabBau2002,STC04,2007ApJ...668..594M} but it has been impossible 
 to  tag them unambiguously with the correct $\ell$, $m$ and $n$.

In this paper we analyse a peak around 220.7 $\mu$Hz that has been studied several times as part of a g-mode 
candidate using different instruments on board SoHO (see for example \citet{Gab99,Fin01,STC04,2007ApJ...668..594M}) 
but also from the theoretical side (see for example \citet{CoxGuz2004} and references therein). To do so, we 
start in section 2 with a brief description of the helioseismic instruments used in this work and we analyse in 
detail the data of the VIRGO/SPM instruments (section 3). In section 4, we look for an instrumental origin
 for this peak without success by analysing all the housekeeping parameters of the VIRGO package as well as 
 the SoHO pointing. Once it is stablished that this peak seems to have a solar origin we check for its
  presence 
 in all the other instruments of the VIRGO package (section 5) and in the velocity instruments GOLF, 
 MDI and GONG (section 6). We then finish by discussing its possible nature.

\section{Instrumentation and data analysis}

The data of VIRGO (Variability of IRradiance and Gravity Oscillation), GOLF (Gravity Oscillation at Low 
Frequencies) and MDI (Michelson Doppler Imager) on board SoHO (Solar and Heliospheric Observatory) satellite have 
been used in this research together with the GONG (Global Oscillation Network Group) ground-based netwok.
 
\subsection{SoHO/VIRGO}

The VIRGO package was designed to study the characteristics of pressure and internal gravity modes by 
observing irradiance and radiance variations, to measure the solar total and spectral irradiance and to quantify 
their variability (Fr\"{o}hlich et al.\ 1995, 1997). It is composed of three different types of sensors:
\begin{itemize}
\item Two types of absolute radiometers (one VIRGO/DIARAD and two VIRGO/PMO6-V) for the measurements of solar 
total irradiance and its variations with high accuracy and precision. The cadences of VIRGO/DIARAD and VIRGO/PMO6-V are 180s and 60s respectively.
\item Two 3-channel sunphotometers (SPM), one  permanently exposed to sun lght and another for backup, set at at 402 nm (blue), 500 nm (green), 
and 862 nm (red), looking at the Sun as a star with a 60 s cadence. 
The bandwidth of the filters is 5~nm.
\item One Luminosity Oscillation Imager (VIRGO/LOI) for the measurements of the radiance in 12 pixels over the 
solar disc. The filter is at 500 nm with a bandwidth of 5 nm.The cadence is 60s.
\end{itemize}

In  1998 June, {\itshape SoHO} was lost
 for several months, but, after a search campaign, was
finally found and resumed  operations around  1998 October. The VIRGO data after  {\itshape SoHO}'s 
``vacations'' show  the same high quality as before the temporary loss of the probe.

\subsection{SoHO/GOLF, SoHO/SOI/MDI and GONG}

\begin{itemize}
\item
SoHO/GOLF is a resonance scattering spectrophotometer (Gabriel et al.\ 1995, 1997)
 that measures the line-of-sight velocity using the sodium
doublet, similar to the IRIS and BiSON ground-based networks. The
GOLF window was  opened in 1996 January and became fully
operative by the end of that month. Over the following months,
occasional malfunctions in its rotating polarizing elements were
noticed that led to the decision to stop them in a predetermined
position; truly non-stop observations began by 1996 mid-April.
Since then, GOLF has been continuously and satisfactorily operating
in a  mode unforeseen before launch, showing fewer
limitations than anticipated. The signal, then, consists of two close
monochromatic photometric measurements in a very narrow band (25
m\AA) on a single wing of the sodium doublet.
This signal has been calibrated into velocity \citep{UlrGar2000,garcia05} and  is indeed  similar in nature
to other known velocity measurements, such as those of IRIS and BiSON 
(Pall\' e et al.\ 1999). The sampling of the
GOLF data used in this paper is 60 s. Before {\itshape SoHO}'s vacations (1998 June),
GOLF data were obtained in the blue wing of the sodium line; thus, after the {\itshape 
SoHO}
vacations the GOLF team decided change to the red wing of the sodium line (see Garc\'\i a et al. 2005, for the latest report on the GOLF instrument).

\item The Solar Oscillations Investigation (SOI) uses a Michelson Doppler Imager (MDI) type of instrument 
\citep{1995SoPh..162..129S}. MDI  consists of a pair of tunable Michelson interferometers,
which image the Sun onto a 1024 pixel x 1024 pixel
CCD camera in five wavelengths across the Ni i 676.8 nm line. These resolved data can be processed by
forming a weighted combination of the pixel signals to yield a
proxy for a Sun-as-a-star response (see \cite{Henney99}).

\item  The ground-based Global Oscillation Network Group (GONG;
Harvey et al. 1996) consists of six sites, with instruments that use
the Fourier tachometer approach to observe the Doppler shift, in
the Ni i line, with 1024 pixel resolution. Here, a
Sun-as-a-star proxy was formed from a simple integration over
all pixels with a cadence of 60s.
\end{itemize}

\section{The 220.7 $\mu$Hz peak seen in VIRGO/SPM}
A long time series of 4098 days of VIRGO/SPM data has been used in this work starting on 1996 April 11.  
As our purpose is to study the time evolution of signals at low frequency, a total of five independent subseries 
of 800 days have been computed. Data were available to allow a 50 day shift up to June 2007. These 66 
overlapped series were used only for plots while non-overlapping data were used as input for the statistical tests we have carried on. 

The slow trends in the time series, due to the degradation in the instruments and long-term solar variability,
 have been removed by applying a running mean filter of one day. To check whether this filter could affect the detected 
 signal, we have also used a backwards difference filter, in which every measured point is substituted by the 
 difference of two consecutive points  $\delta f_{n}=f_{n+1}-f_{n}$. To recover the correct amplitudes in the 
 power spectrum, this latter should be divided by the transfer function of the filter $Q(\nu)$, defined as follows \citep{GarBal08}: 
\begin{equation}
Q(\nu)=[2sin(\pi\nu\Delta t)]^2
\end{equation}
 $\nu$ being the frequency and $\Delta t$ the sampling of the data.

Both filters gave the same results and, in the rest of this paper, we work only with the time series filtered by the 1-day running mean.

We therefore computed several power density spectra using a Fast Fourier Transform (FFT) algorithm and
 built the time-evolution power diagrams used in this work. Each of them has been computed from the 
 time series which have been extended by four equal time intervals of zero signal . This oversampling makes it easier to detect the bins in which the 
 power is concentrated \citep{GabBau2002}. We  also verified that a sine wave fit (SWF), computed 
 in steps of 0.0001 $\mu$Hz between 220.5 and 221 $muHz$, yields the same results. Therefore, we have used the 
 normal zero-padded FFT as it is much faster than the SWF. 

The time-evolution power diagrams are built as follows: The 66 power spectra of the overlapped time series are 
computed and plotted vertically using  a colour scale for the power. The vertical axis is the frequency of the 
power spectra with the colour equivalent to the power as indicated on the right-hand side of the diagram and the 
horizontal axis the number of the time series from 0 to 65, i.e.\  the time span corresponding to the 
time series. Looking these time-evolution power diagrams we  know  at which frequency and for how long 
 a signal can have enough power to be observed above the noise level.
   
\begin{figure}[!htb]
\centering
\begin{tabular}{c}
	\includegraphics[width=13pc,angle =90]{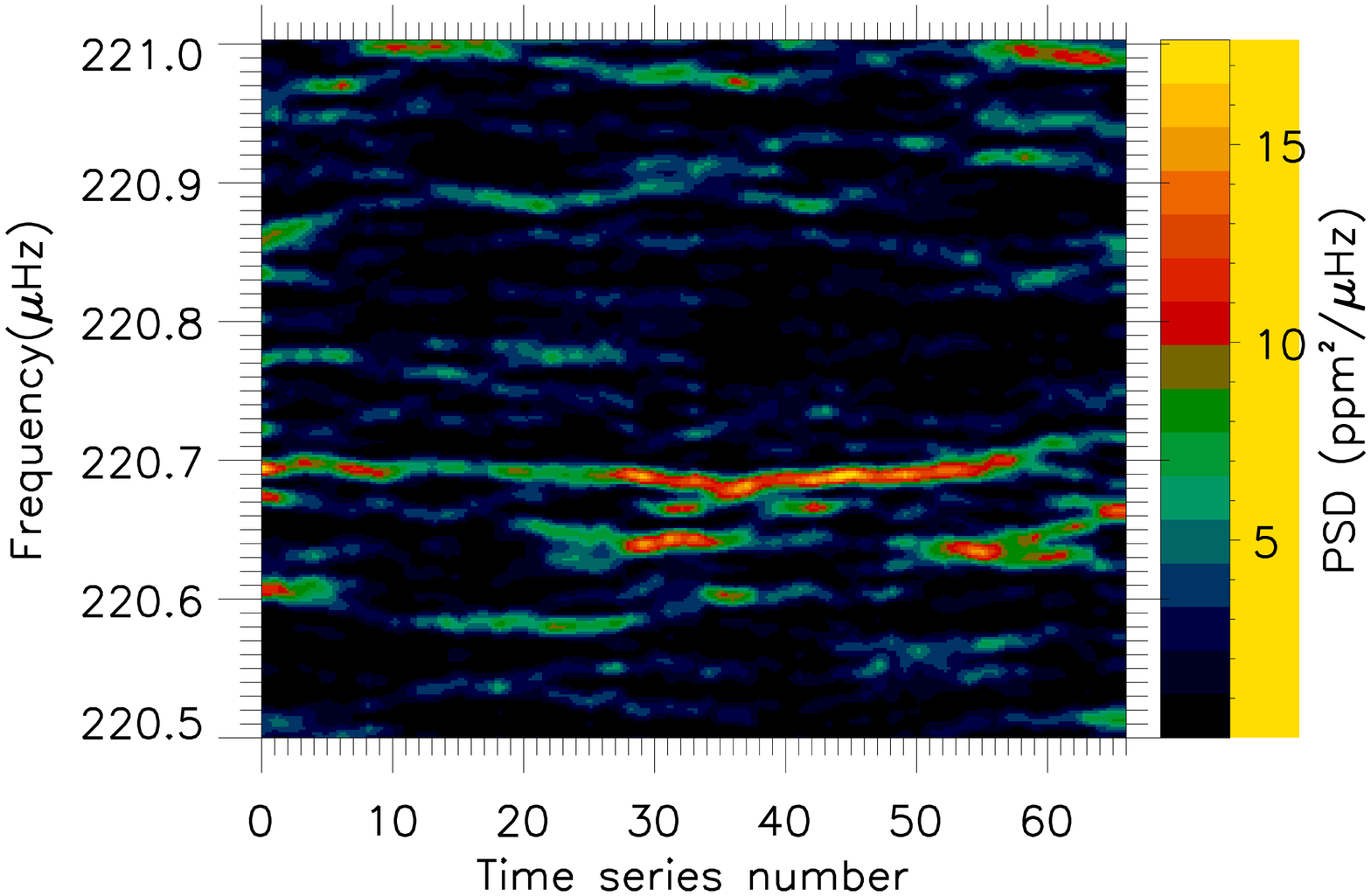} \\
  	\includegraphics[width=13pc,angle =90]{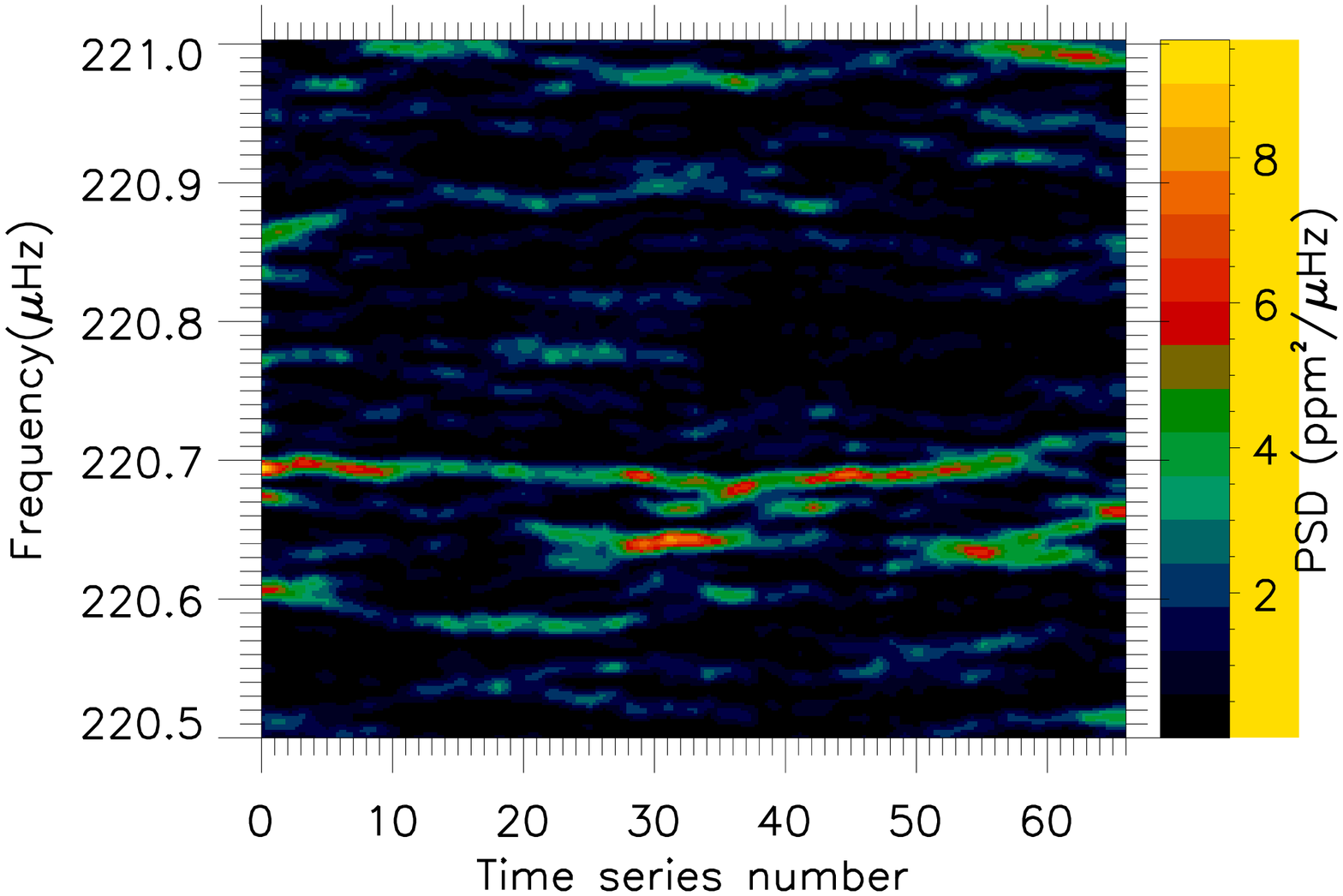} \\
	\includegraphics[width=13pc,angle =90]{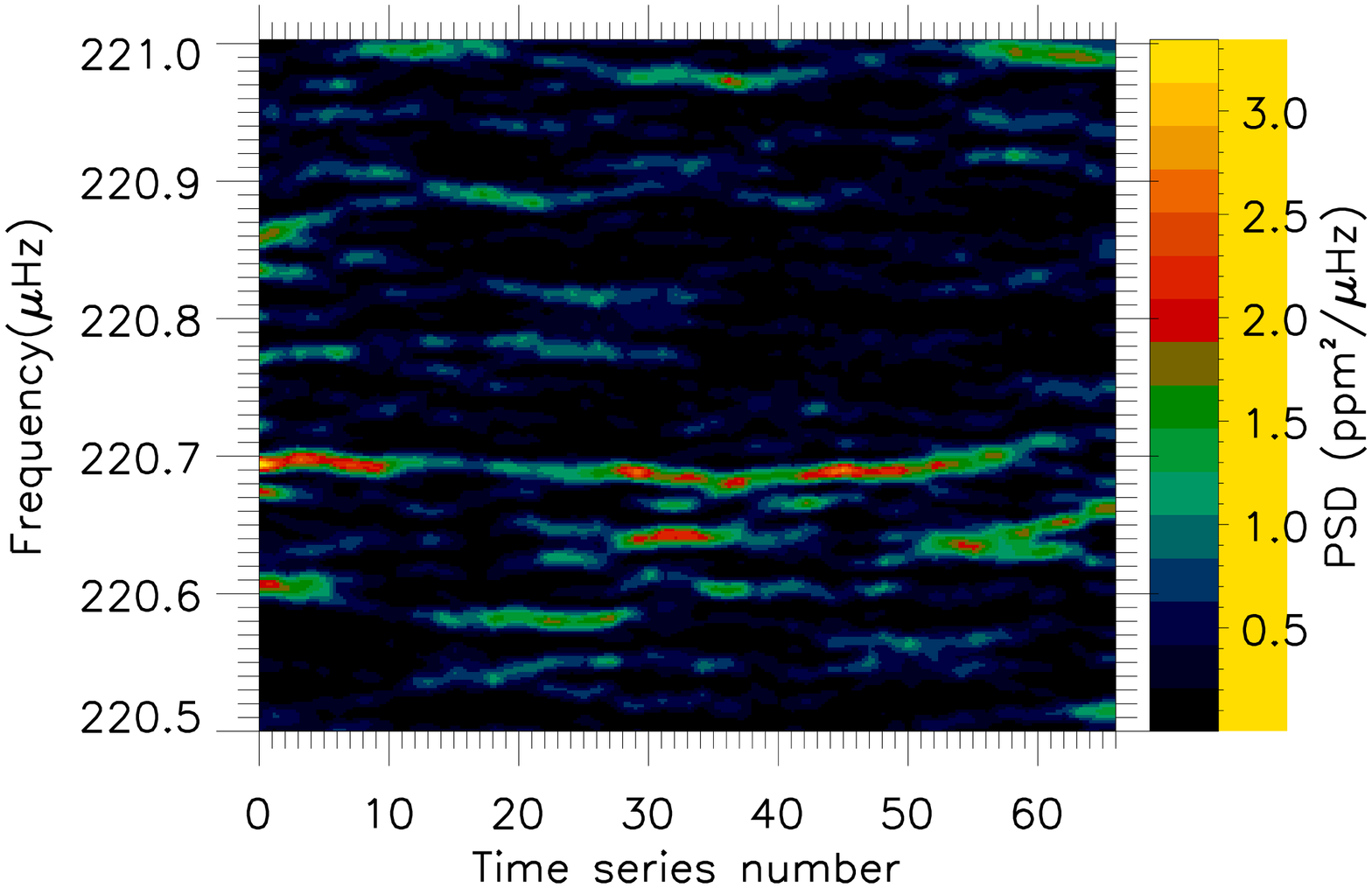} \\
\end{tabular}
\caption{\label{spms}VIRGO/SPM time-evolution power diagram of channels blue, green and red (top to bottom 
respectively) from April 1996 to June 2007.  A stable signal at  220.7 $\mu$Hz  is clearly observed. }
\end{figure}

 In figure~\ref{spms} the time-evolution of the three VIRGO/SPM channels are shown for the frequency range
  220.5--221.0 $\mu$Hz.  The x-axis spans 11 years of the SoHO mission.  A clear signal is observed 
  in the blue channel (top) at around 220.7 $\mu$Hz, which is stable in time with power that goes from 6--7  to 
  16--17 $ppm^2/\mu$Hz. Around time series 60 this signal seems to  change its frequency slightly by around 0.3 
  $\mu$Hz. In the green and red channels the same continuous signal is visible as in the blue one but with 
  the expected decrease in power with wavelength. It is also important to note that in all these VIRGO/SPM channels 
  a second high-amplitude signal is visible at $\sim$ 220.64 $\mu$Hz, parallel to the previous one, from time 
  series 20 until the last one but with a small gap between time series 44 and 48.

\subsection{Confidence levels and Monte-Carlo simulations}

\begin{figure}[!htb]
\includegraphics[width=0.36\textwidth,angle =90]{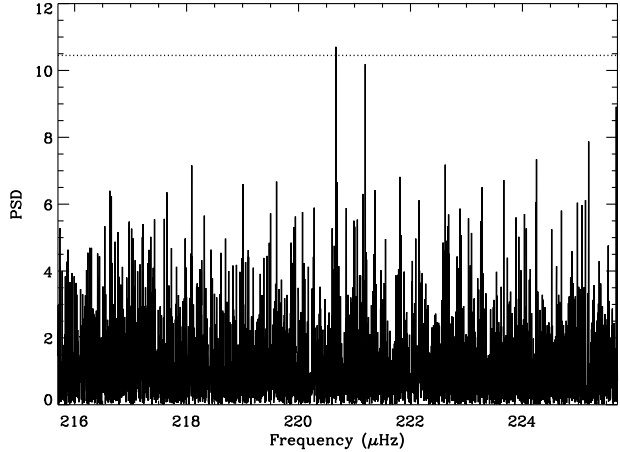}
\caption{\label{full}VIRGO/SPM blue channel power spectrum density computed with 4098-day time 
series starting on 1996 April 11. The horizontal dotted line corresponds to the 90\% confidence level 
that a peak above this line would not be due to noise.  }
\end{figure}

In the previous section we saw that there is a persistent signal around the target frequency of
 220.7 $\mu$Hz. Indeed, the VIRGO/SPM blue channel power density spectrum of the full length time series 
 (see Figure~\ref{full}) shows the presence of a peak at a precise frequency of 220.667 $\mu$Hz above the 
 90$\%$ confidence level computed in a   10 $\mu$Hz window following \cite{App00}.

Using subseries of 800 days and a frequency window of 10 $\mu$Hz, the power level above which an observed
 peak has a 90\% probability not due to noise is 8.87$\sigma$ (e.g. see \cite{App00}). In the case of 
 zero-padded data, the points are no longer independent and  are correlated. Therefore, Monte-Carlo 
 simulations should be used to derive a correction for the above-mentioned confidence level.  In our case, 
 for a padding factor of 5 we have added a correction of $\mathrm{ln}(2.8)=1.03$ derived by \cite{GabBau2002} 
 to the threshold computed using non-zero-padded data. In this conditions the 90\% confidence level at around 
 9.9$\sigma$. Using the VIRGO/SPM blue channel, we found that the peak we are studying has a maximum power in a
  range between 8.74 and 10.4$\sigma$ considering only five independent realizations of 800 days. In Figure~\ref{spms} 
  the 90\% limit is obtained at around 14 $ppm^2/\mu$Hz (orange colour in Figure~\ref{spms}). This means that, for example,
   most subseries between the 28th and the 58th have the peak above the 90\% confidence level, as well as other subseries 
   such as those at the very beginning of the time-span. It is important to notice that in this case the $\sigma$ has 
   been averaged over the 66 time series and the value of 14 $ppm^2/\mu$Hz is an averaged magnitude. 

We are interested in knowing  the probability of having a signal with the same properties to those that we have found in 
the VIRGO/SPM blue channel; i.e.\ a peak that is above the 90$\%$ level in the full power density spectrum of more 
than 4098 days, and that is also present in the five independent subseries of 800 days with similar levels to what we have with 
this instrument (i.e.\ not necessary all above a 90\% confidence level in these individual small subseries but around that 
level). This latest condition would be much more restrictive because  it means that the peak should maintain a certain 
coherence during the full time-span. A Monte Carlo simulation of 1 million iterations has been done by simulating Gaussian 
noise time series of 4000 days that have been cut into five intervals of 800 days. To speed up the procedure we have not computed 
the full spectrum of the 4000 days but only the average of the power density spectrum of the five independent realizations of 800 
days (which have an SNR of $\sim$ 9.2 $\sigma$ in the VIRGO/SPM blue channel). Thus, the algorithm looks first for a signal 
in the average spectrum with $\sim$0.9 times the level found in the VIRGO/SPM blue channel (8.3$\sigma$) and, if it is 
found, it looks for the presence of that signal in the five subseries (again with levels of 0.9 times those of VIRGO).
 Any signal with these properties found in the 10 $\mu$Hz window will be flagged as a positive identification. The results 
 show that, in a window of 10 $\mu$Hz, the 220.7 $\mu$Hz signal has a likelihood of 99.8$\%$ (which is reduced to 91.3\% if 
 we only consider the constraint on the averaged spectrum). We have also checked how the likelihood is degraded when a bigger 
 window is considered. Thus, for the 20 and 30 $\mu$Hz windows we obtain  99.6 and 99.4\% respectively.

We can conclude that it is extremely difficult to find a pure noise signal above the 90\% confidence level after $\sim$4000 days 
and with a coherence with time as  found in the VIRGO/SPM instruments.

 \section{Possible instrumental origin of the signal inside VIRGO and SoHO}
 
Once this interesting signal has been detected in VIRGO/SPM the main question is to investigate its origin; in other words, determine 
if it is of solar or instrumental origin. In this section, we study all the possible non-solar 
origins of this signal, from the orbital and pointing corrections of the spacecraft to the housekeeping parameters 
(hereafter HK) of the VIRGO package. 

Periodic manoeuvring of the SoHO probe at this frequency (220.7 $\mu$Hz, i.e. a period around 1.25 hours) due to orbital 
adjustments or pointing corrections could modulate the signal of the instruments on board as a tracking system can produce guided frequencies.
On the other hand, a temperature variation at this frequency could also modulate the observed signal. These temperature 
variations could originate in the sensor itself or in other instrument subsystems.

For all these parameters we follow the same analysis we as for the VIRGO data; thus, we build the corresponding 
time-evolution power diagrams and we compare them with the VIRGO/SPM ones. If the 220.7 $\mu$Hz signal is produced by the 
temporal variations of some of these parameters, the time-evolution power diagrams of both VIRGO/SPM and the parameter must 
be highly correlated.

For this purpose we analyse in the following subsections orbital and pointing corrections of the SoHO spacecraft and the different 
HK parameters of the VIRGO package that might modulate the signal. It is important to note that some of the HK data 
are in the scientific telemetry of VIRGO and have a cadence of 60s, while others are in the HK telemetry and have a cadence of 180s.

\subsection {Orbital corrections}

The radial distance is reduced to 1AU by the usual quadratic law $S_{0}=S \cdot r^2$,  $r$ being the spacecraft-to-Sun distance in 
astronomical units. This correction normalizes the spectral irradiance to the solar constant definition and removes signal modulations 
due to movements of the Earth, Moon and planets in their orbits.

The observed radiation $S$ of a moving blackbody source is 
\begin{equation}
S=S_{0}  \frac{(1-v)^2}{(1-v^2)} 
\end{equation}
where $S_{0}$ is the radiation in motionless conditions and
 v is the speed in units of the light speed, c. With SoHO velocity being a few $10^{-6}$ of the speed-of-light one can safely omit terms in 
 $v^2$ and thus approximate the reciprocal formula
\begin{equation}
S=S_{0} \frac{(1-v)^2}{(1-v^2)}\sim \frac{S_{0}}{(1-2v)} \sim S_{0} (1+2*v)
\end{equation}
This Doppler correction removes a tiny ($10^{-5}$), slow (Halo orbit period is 6 months) modulation of the measured irradiance.

In this way, the orbital correction applied to the three channels of VIRGO/SPM is:
\begin{equation}
SPM_{channel}=SPM_{channel} \cdot radius^2 \cdot (1+2 \cdot vel)
\end{equation}
where ``radius" is the spacecraft-to-Sun distance in astronomical units and ``vel" is the radial velocity in units of the speed-of-light.

\begin{figure}[!htb]
\centerline{%
\begin{tabular}{c@{\hspace{1pc}}c}
\includegraphics[width=13pc,angle =90]{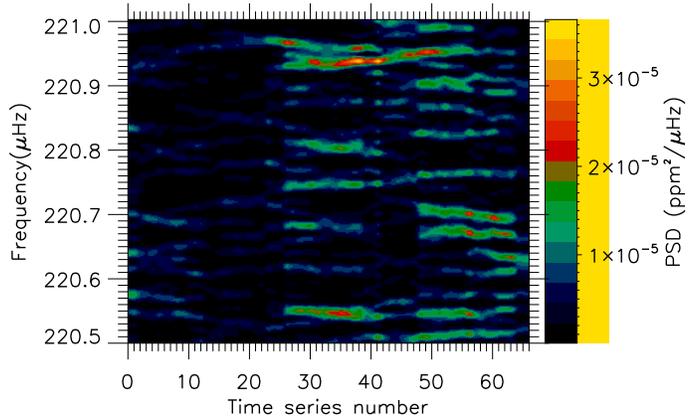}
\end{tabular}}
\caption{\label{orbit}Time-evolution power diagram of the orbital corrections applied to the VIRGO data. In addition to the order of
 magnitude which is 6 orders of magnitude smaller than VIRGO/SPM no correlation is found.}
\end{figure} 

The orbital parameters (radius and vel) are provided by NASA in a 10 minute  cadence and  are linearly interpolated to get the 
same 60s as VIRGO/SPM. The time series of the orbital correction applied, i.e.\ $radius^2(1+2\cdot vel)$, has been analysed in the same 
way as VIRGO/SPM and the resulting time-evolution power diagram diagram is shown in Figure~\ref{orbit}.  The orbital correction 
signal is around 6 orders of magnitude smaller than the VIRGO/SPM one and no correlation has been found.

\subsection{Spacecraft Pointing}

\begin{figure}[]
\centering
\begin{tabular}{c}
	\includegraphics[width=12pc,angle =90]{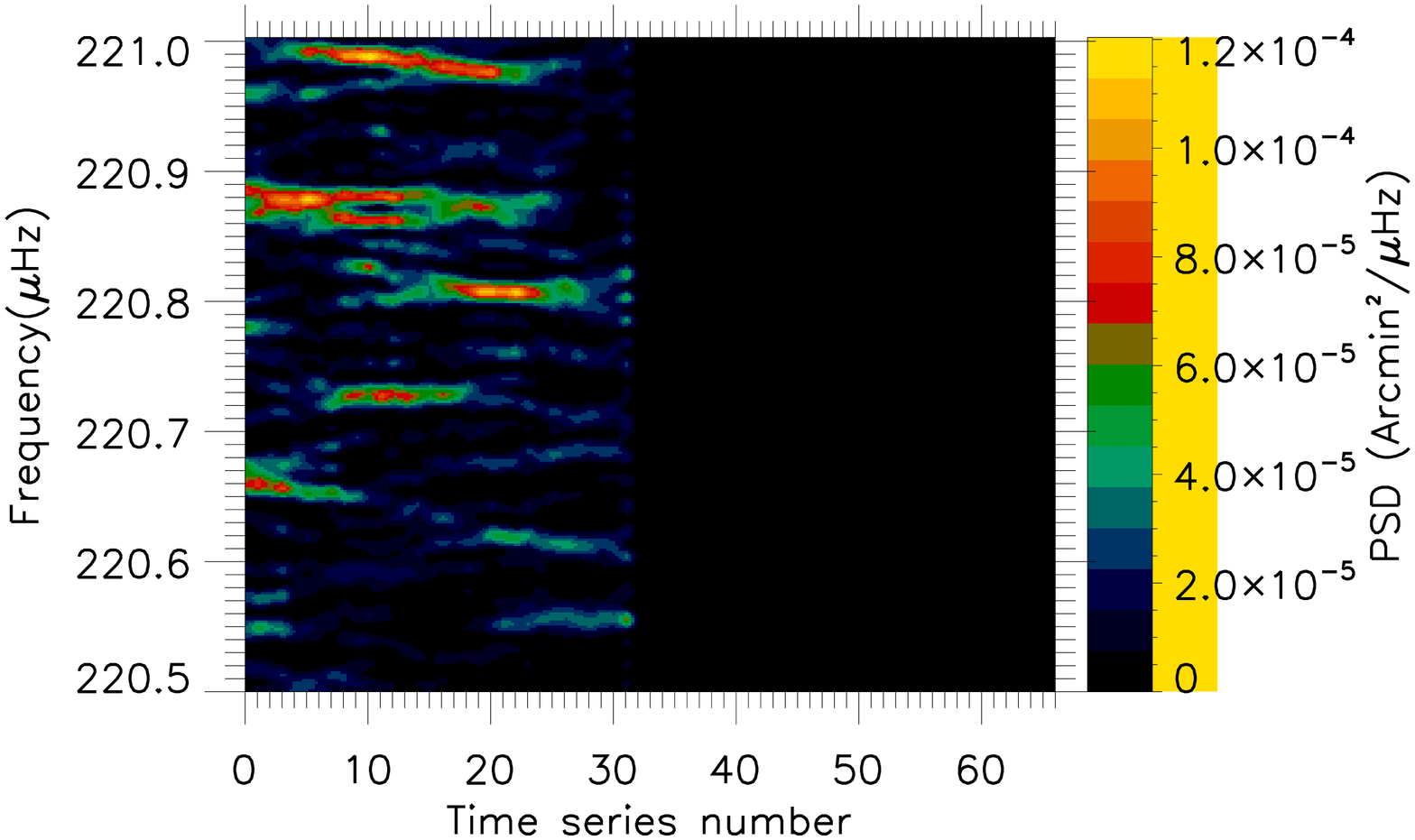} \\
  	\includegraphics[width=12pc,angle =90]{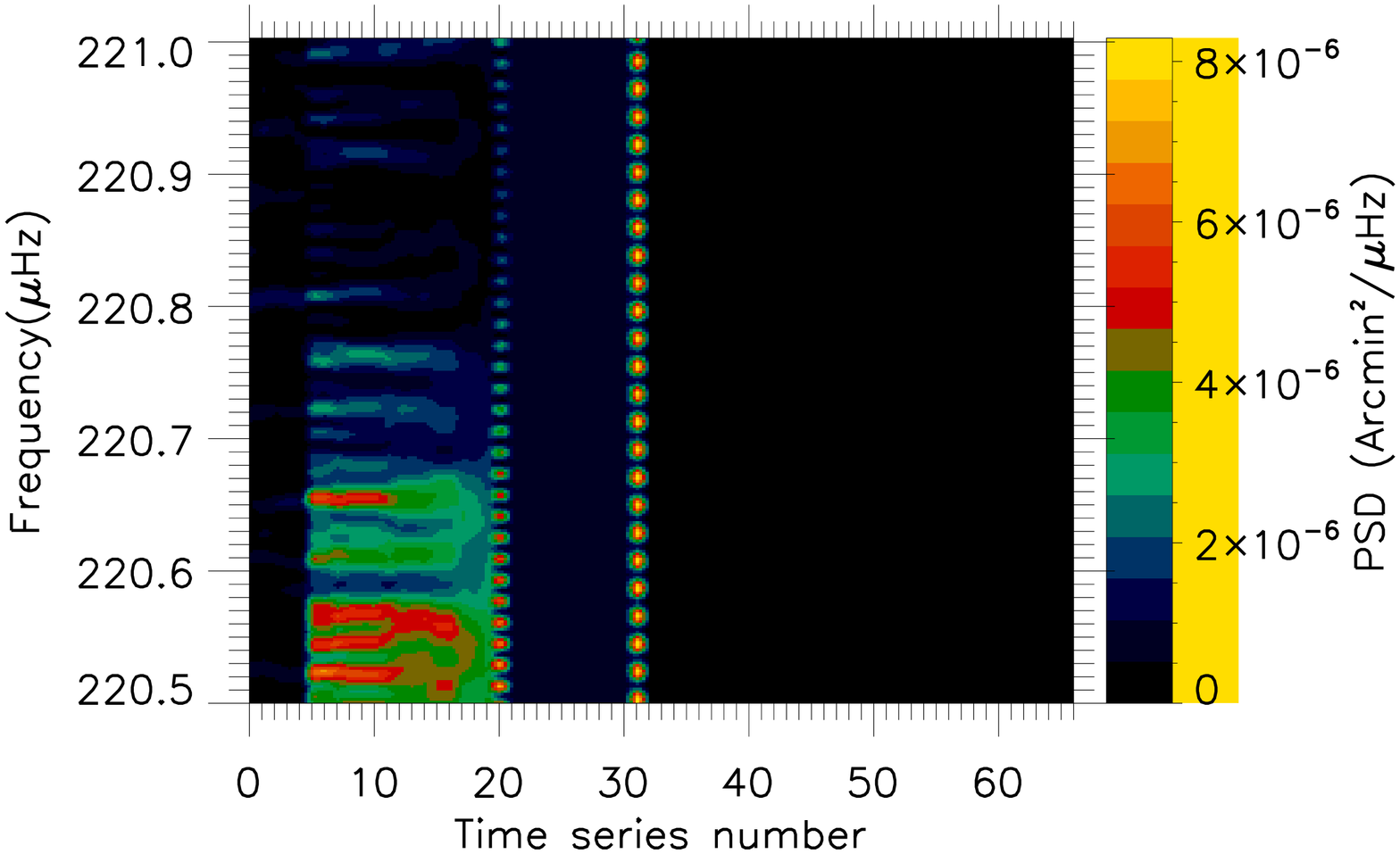} \\
	\includegraphics[width=12pc,angle =90]{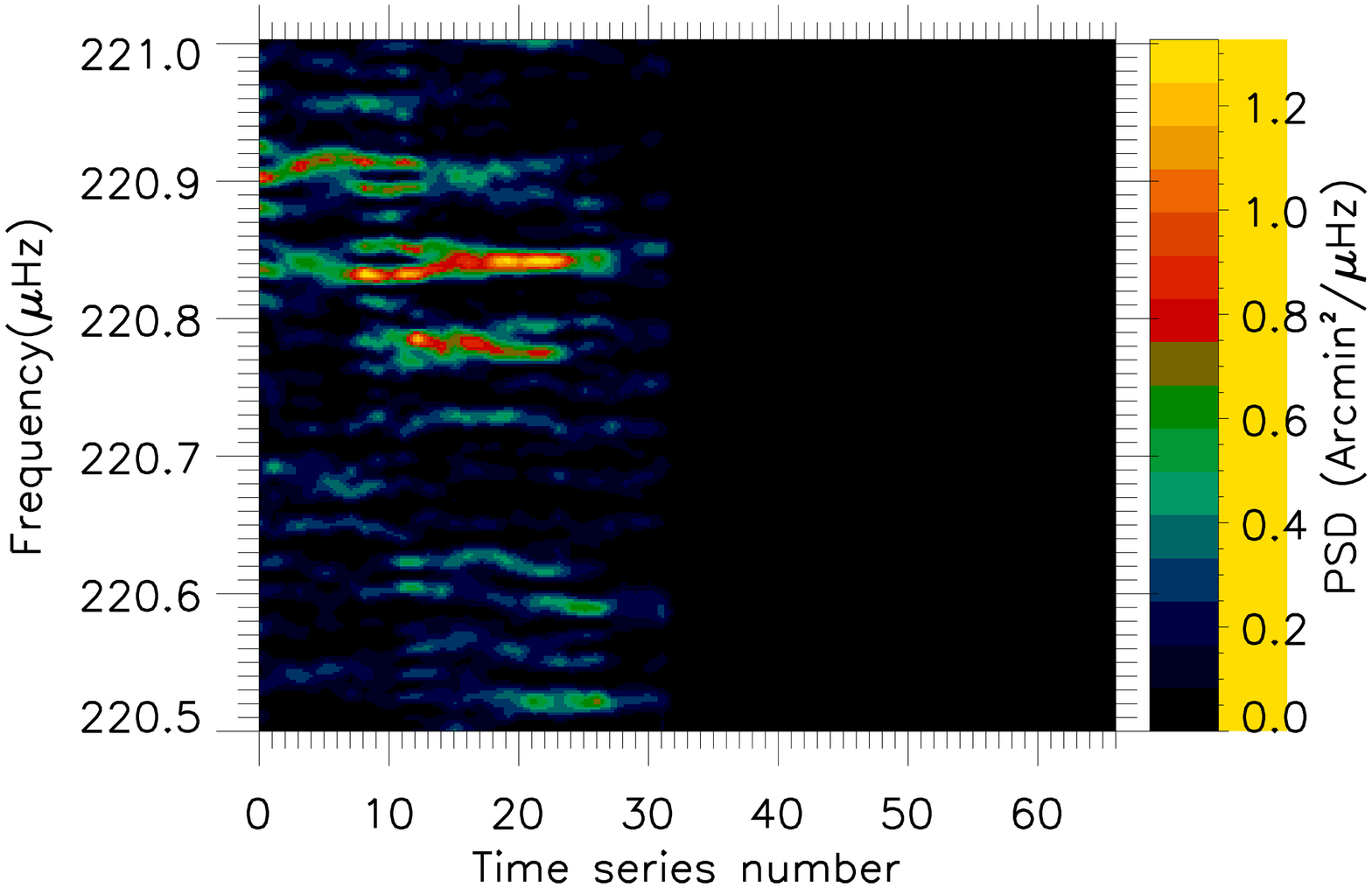} \\
\end{tabular}
\caption{\label{pointing}Power time-evolution diagrams of the pitch, yaw and roll angles (respectively from top to bottom) 
of the SoHO spacecraft. Only data up to September 2002 (time series 32) are available. No correlation with the first 32 time series of VIRGO/SPM is found.}
\end{figure} 

The three critical flight dynamics parameters are rotations in three dimensions around the vehicle's coordinate-system origin, the 
centre of mass. These angles are pitch, roll and yaw. Pitch is the rotation around the lateral or transverse axis. Therefore, 
movements of the spacecraft to the north or south of the Sun. Yaw is the rotation about the vertical axis; thus, movements of 
the spacecraft to the west or east of the Sun and, finally, roll is a rotation around the longitudinal axis i.e.\ movements of the 
spacecraft from the north or south to the west or east of the Sun.

For an instrument that looks at the Sun as a star (integrated light) the most plain pointing correction would be divided 
by $cos(\sqrt(yaw^2 + pitch^2)$, i.e.\ the cosine of the angle between instrument optical axis and the line-of-sight
direction. Nevertheless, this correction was never applied to VIRGO/SPM because the correction would have been negligible. 
Also, in December 2001, NASA  discontinued the CDHF (Central Data Handling Facility), which was the facility in charge of processing, 
producing and distributing the SoHO telemetry and the ancillary data products. The production of all these data were continued in
 others ways but the production of attitude data was stopped in September 2002. Indeed, when SoHO is in normal mode, the attitude
  follows nominal attitude well enough for most purposes, and because the roll determination had large errors (because of certain procedural problems).

Even knowing that, it would be very unlikely that the pointing maneuvres could  modulate any signal in the SoHO instruments; the three 
angles have been analysed and their time-evolution diagrams computed (see Figure~\ref{pointing}).
 
The available attitude data concerning pitch, yaw and roll angles were obtained from the NASA archive from 1996 April 11 to 2002 September
 22 and we built the time series of the three angles. These data have a cadence of 10 minutes and have been used with this sampling rate 
 because it is good enough for our purposes. The length of these time series enables us to get 32 time series of 800 days (each shifted 50 days 
 with respect to the previous one). This is approximately half of the time-evolution power diagrams used in the VIRGO/SPM. This length is 
 sufficient to see if any correlation exists between pointing and SPM signals during the common period (around 6 years).

The pitch angle has a very constant value of around -3.3 arcmin during the time span, with spikes of 5 arcmin and only a few of them 
with higher values, between 6 and 13 arcmin. These latter  are probably due to spacecraft maneuvres. The associated time-evolution 
diagram is shown in Figure~\ref{pointing} ({\it top}). Some power density has been found at a level of $10^{-5}  (arcmin)^2$/$\mu$Hz with
 no visible correlation with the VIRGO/SPM signal.

The yaw angle is zero during practically the whole  time-span considered with some spikes around 1.7 arcmin and only a few between 6.8 to 
12 arcmin. This yields  a pure noise time-evolution diagram (see Figure~\ref{pointing} ({\it medium})) with a power density of around $10^{-6}  
(arcmin)^2$/$\mu$Hz with also no visible correlation with the VIRGO/SPM signal.

Finally, as we have already said, the roll angle does not affect the data achieved by instruments that observe the Sun as a star (integrated 
light) but, in any case, it has also been analysed. The roll angle changes following the Earth orbit between 7.16 and -7.16 degrees with some 
large spikes that have been removed (the roll angle sometimes has large errors) and the time-evolution diagram is shown in 
Figure~\ref{pointing} ({\it bottom}). The power density is around $1 (arcmin)^2$/$\mu$Hz and, once again, no correlation with the measurements 
of VIRGO/SPM has been found.

\begin{figure*}[!htb]
\centering
\begin{tabular}{cc}
	\includegraphics[width=12pc,angle =90]{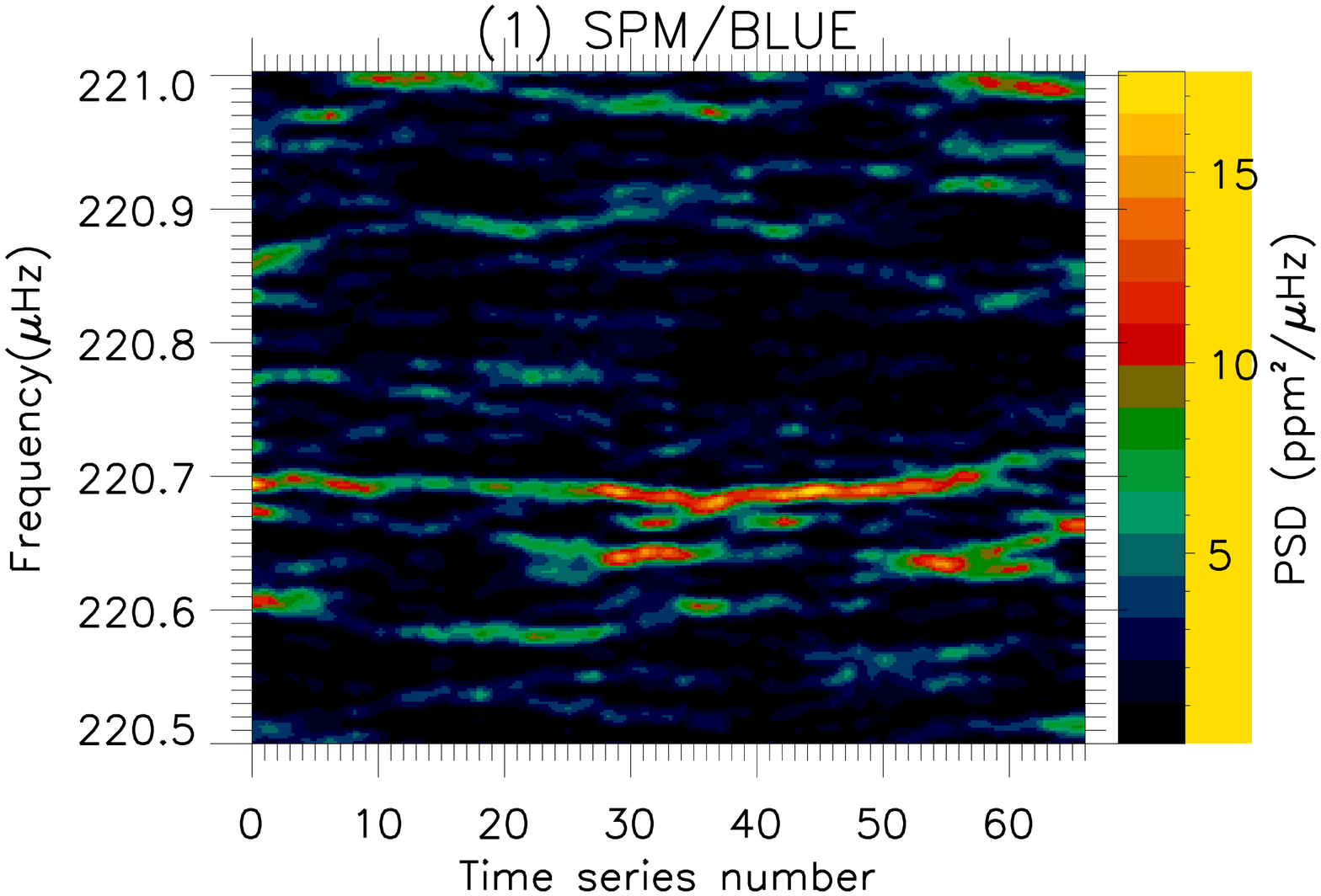} 
	\includegraphics[width=12pc,angle =90]{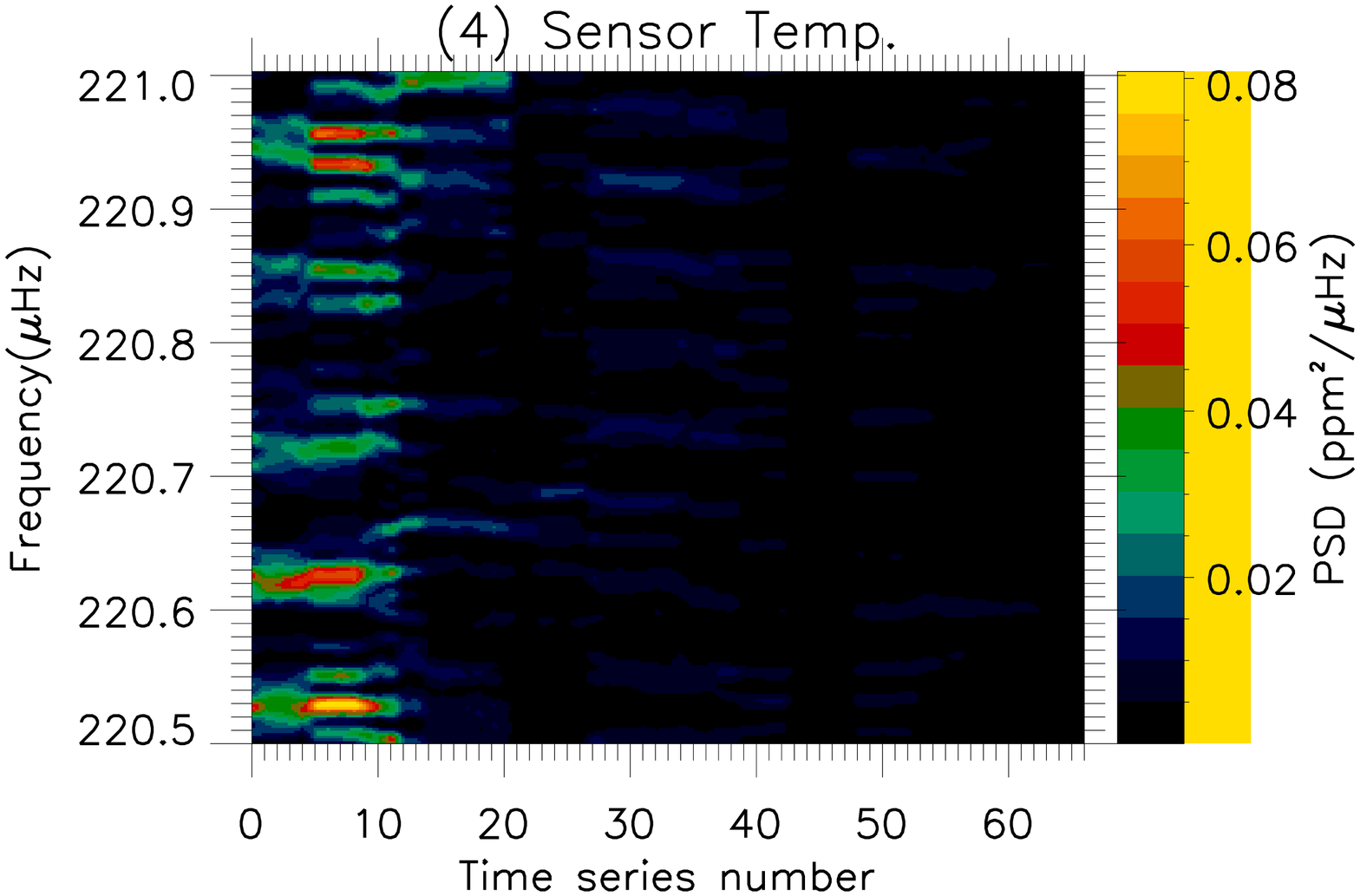} \\
  	\includegraphics[width=12pc,angle =90]{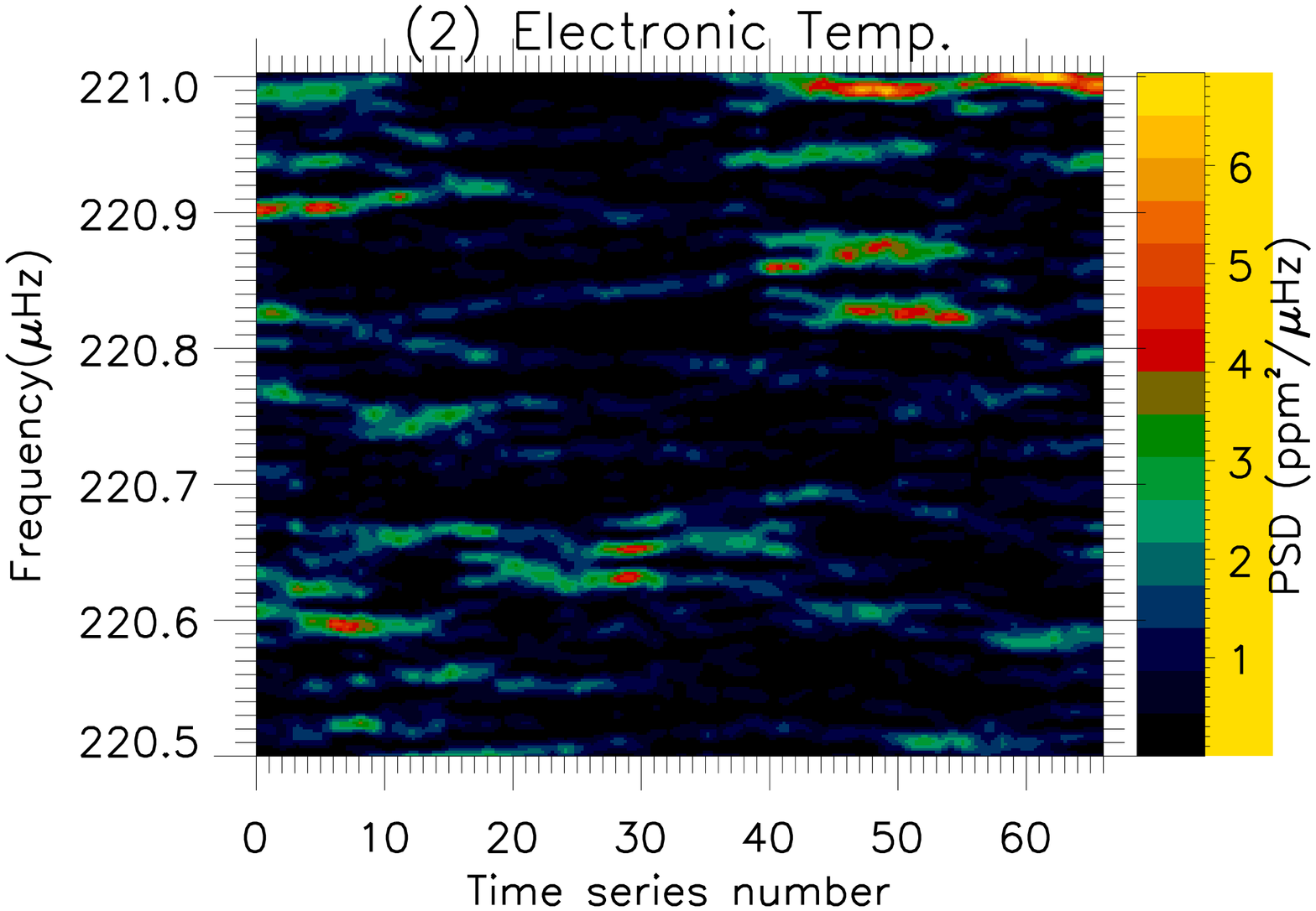} 
	\includegraphics[width=12pc,angle =90]{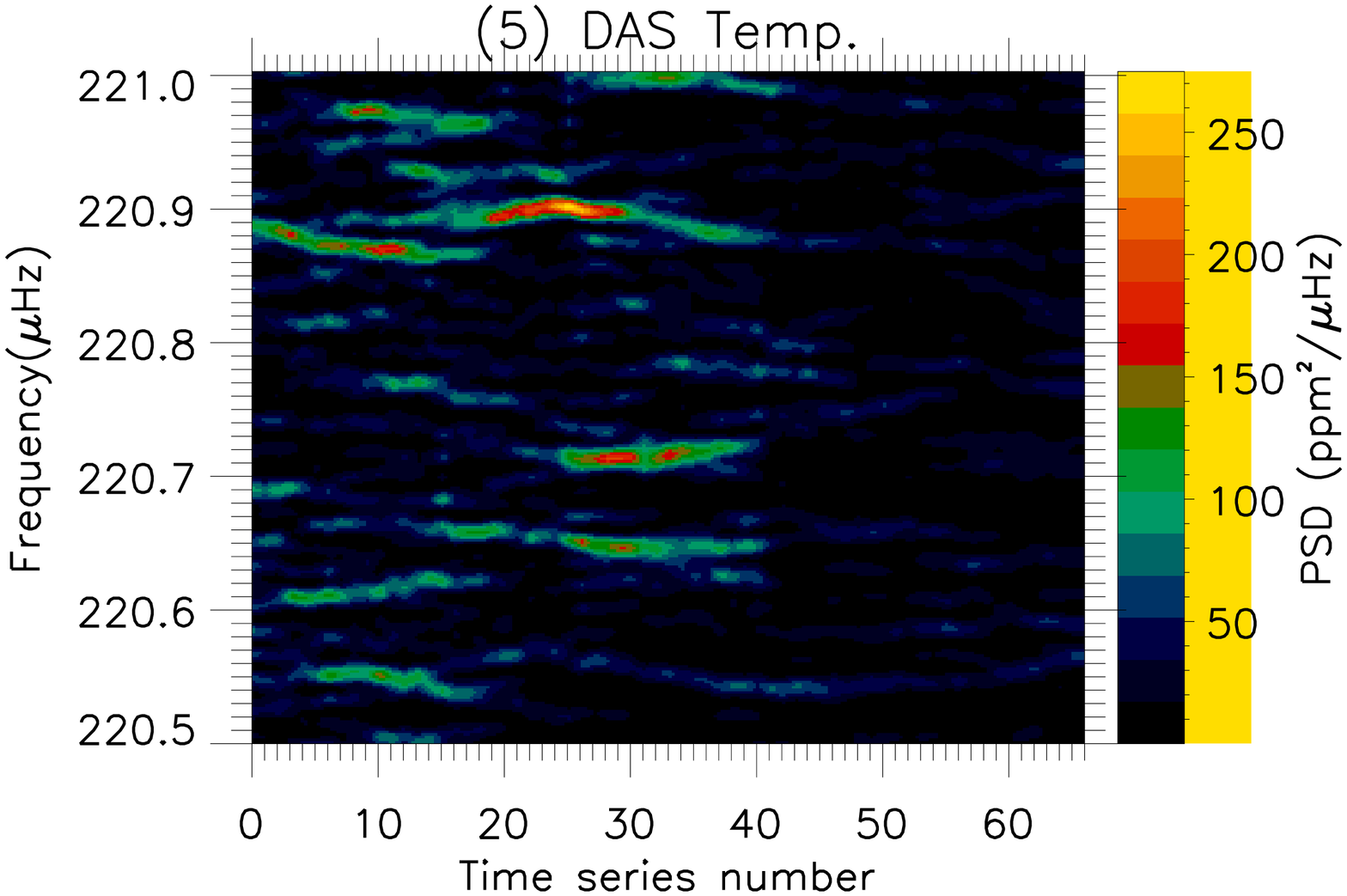}\\ 
	\includegraphics[width=12pc,angle =90]{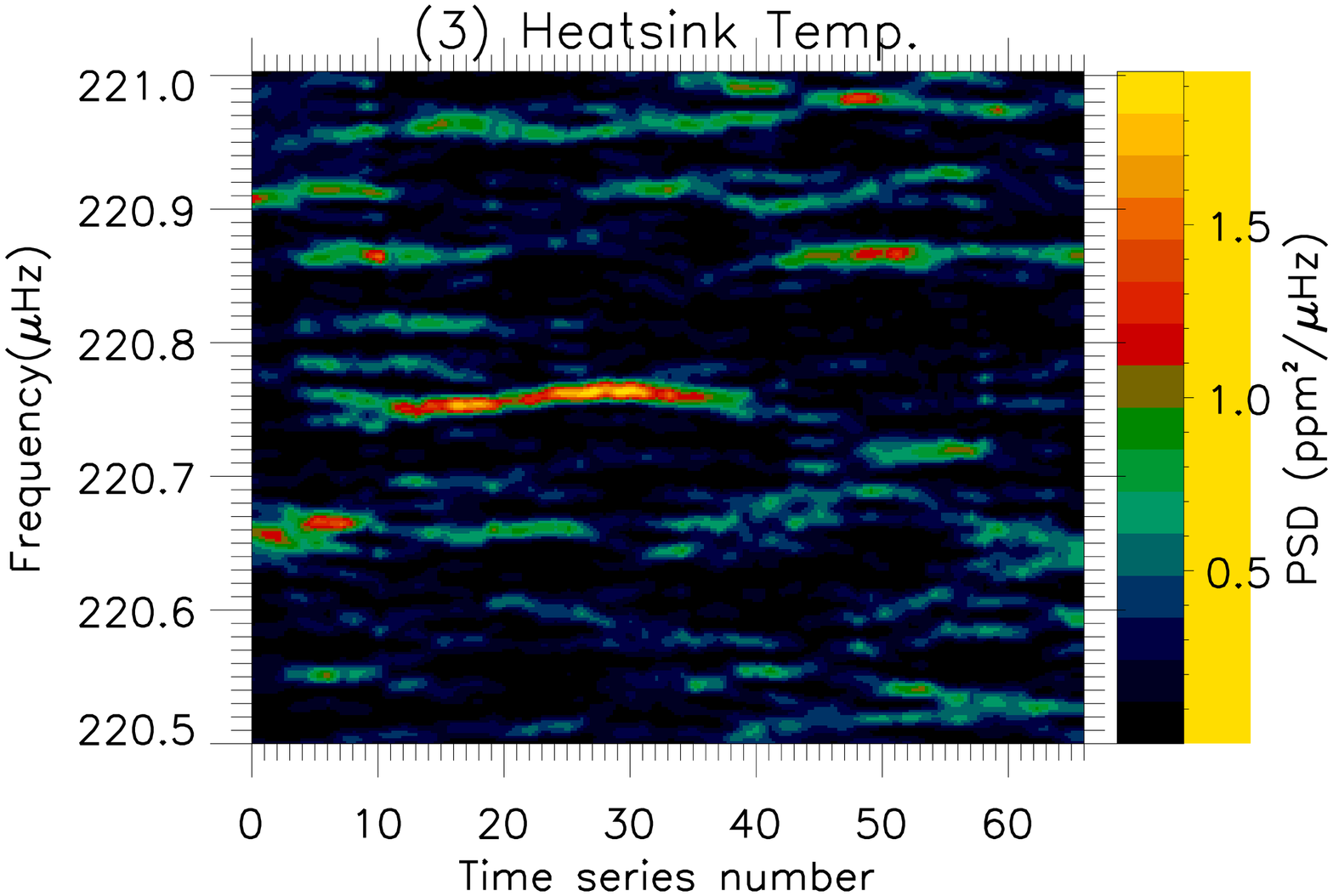} 
  	\includegraphics[width=12pc,angle =90]{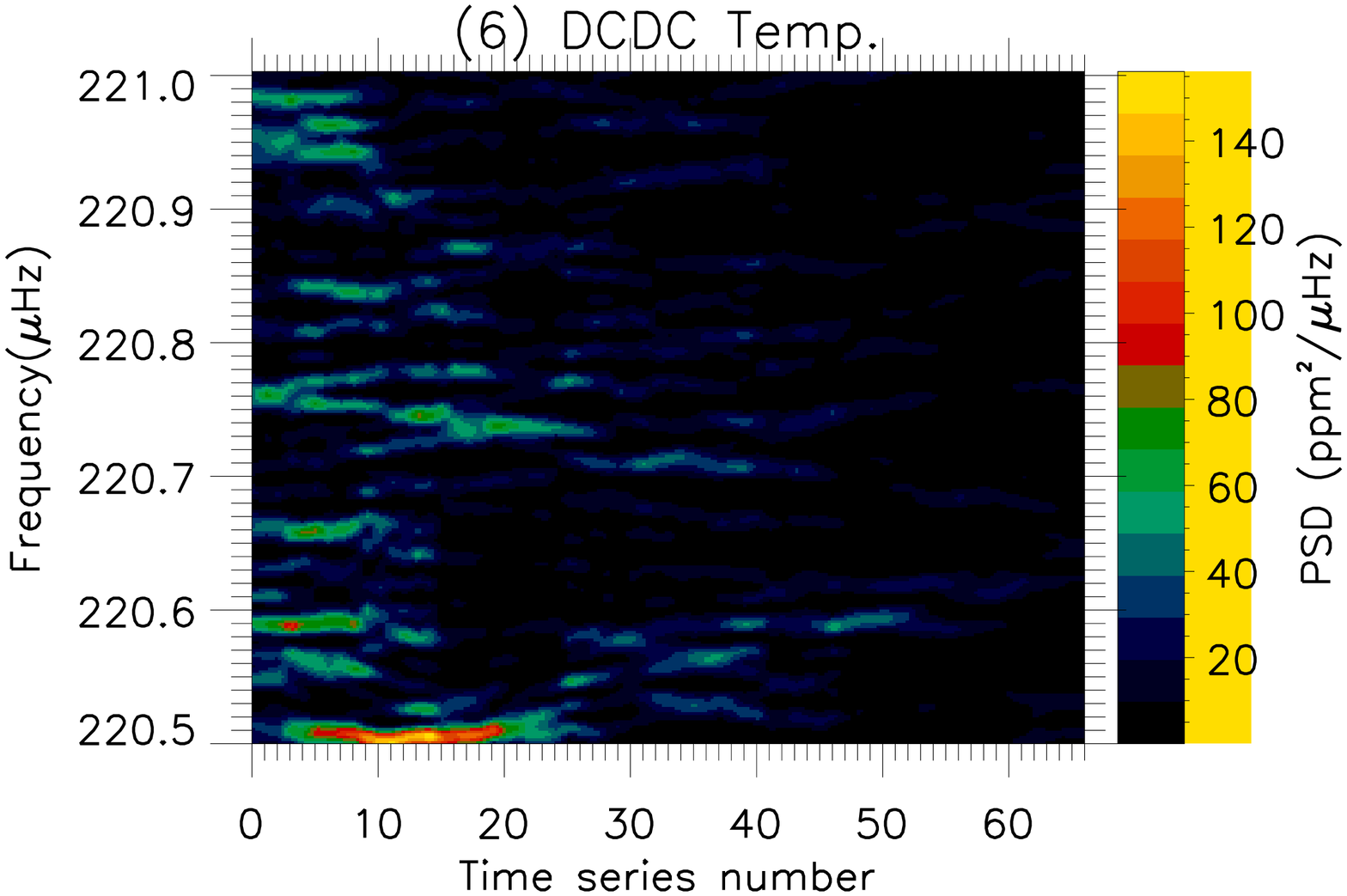}\\ 
\end{tabular}
\caption{\label{Temps}Time-evolution power diagram of the HK temperatures of the VIRGO/SPM package. To simplify the 
comparison we have repeated, on the same scale, the time-evolution diagram of the VIRGO/SPM Blue channel.}
\end{figure*} 

\subsection{VIRGO/SPM temperatures}

\begin {itemize}
\item{VIRGO/SPM sensor temperatures}

The most important VIRGO/SPM temperature is the temperature sensor. Each of the three VIRGO/SPM channels (blue, green and red) are 
corrected by a quantity proportional to each of the temperature sensors (sensor blue, green and red). This correction is applied in 
the level 1 software, so the data we are handling are already multiplied by this quantity.  This correction is:
 \begin{equation}
SPM_{channel}=(1+C_{channel}(TS_{channel}-293.15))
\end{equation}
 where ``channel'' means blue, green or red; $C_{channel}$ is a  constant for each channel and $TS_{channel}$ is the temperature of each of the three sensors.

In Figure~\ref{Temps} the time-evolution power diagram  of the temperature sensor of the blue channel is shown (top of the right column) . 
The fluctuation of this temperature is two orders of magnitudes smaller that the VIRGO/SPM signals at frequencies around 220$\mu$Hz and no 
clear correlation  with VIRGO/SPM signals is visible.

\item{VIRGO/SPM electronic temperature}

The temperature of the SPM electronics has been also analysed to see if there exists some modulation that could produce a periodic variation 
in the output voltage of the low-noise electrometer amplifiers (or in the input current). If this exists, a modulation would go to the 
Voltage Frequency Converters (VFC) of the Data Acquisition System (DAS) and could produce a modulation in the output signal.  

The time-evolution power diagram for the VIRGO/SPM electronic temperature is shown in Figure~\ref{Temps} (middle of the left column). 
The power density is of the same order as in the VIRGO/SPM channels but no correlation with the signal at 220.7 $\mu$Hz has been found.

\item{Data Acquisition System (DAS) temperature}

The Data Acquisition System (DAS) of VIRGO comprises the Onboard Data Handling System (interface for telemetry, telecommands and timing 
signals), multiplexers, Voltage Frequency Converters (VFC) and counters. If the DAS temperature, the VFC or the counters have a periodic 
behaviour, the output number of counts could contain that periodicity. The DAS temperature time-evolution diagram is shown in Figure~\ref{Temps} 
(middle of right column). The power density is an order of magnitude higher than the SPM signal  but again there is no correlation with the signal at 220.7 $\mu$Hz .

\item{VIRGO/SPM Heatsink and DC/DC temperatures}
 
The temperature variations of the VIRGO Heatsink and the VIRGO Power Supply (DC/DC) have been also analysed for security. The Heatsink 
time-evolution diagram (Figure~\ref{Temps})(bottom of the left column) is ten times smaller than the VIRGO/SPM and that corresponding 
to the DC/DC is ten times larger (Figure~\ref{Temps} ,bottom of the right column). In both cases no correlation is found with the 220.7 $\mu$Hz signal.

\begin{figure*}[!hptb]
\centerline{%
\begin{tabular}{c@{\hspace{1pc}}c}
\includegraphics[width=35pc,angle =0]{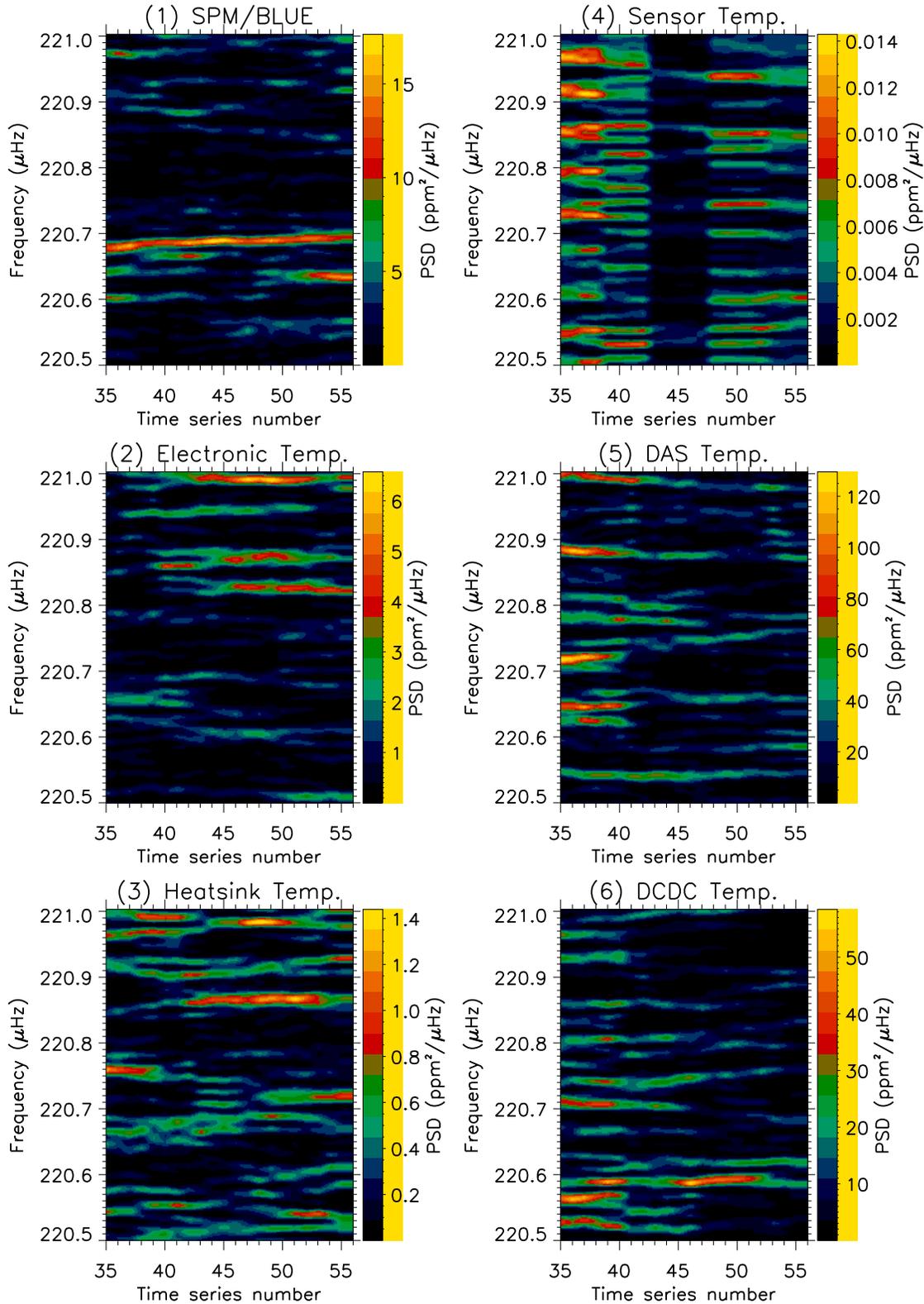}
\end{tabular}}
\caption{\label{zoom} Zoom of the time-evolution power diagrams of the VIRGO/SPM Blue and the VIRGO housekeeping analysed in this research 
for the time series 35 to 55 where the 220.7 $\mu$Hz has higher amplitudes. This zoom helps to clarify the darker parts of some time-evolution 
diagrams produced by the colour scales. None of these temperatures can explain the observed signal at ~220.7 $\mu$Hz .}
\end{figure*} 

Finally, if a signal is the result of a certain temperature modulation, the temperature variation would be higher just where the power of 
the signal is higher. In Figure~\ref{zoom} the SPM/Blue and VIRGO HK time-evolution power diagrams are plotted together but only between time 
series 35 to 55, in which the power of the  220.7 $\mu$Hz signal is stronger in the VIRGO/SPM data. This zoom helps us to see the darker parts of 
some HK time-evolution diagrams produced by the colour scales. None of the temperatures analysed in this section can explain the observed signal 
at 220.7 $\mu$Hz.

\end{itemize}

\section{The 220.7 $\mu$Hz signal in the others VIRGO instruments}

As  was mentioned in section 2.1 the VIRGO package comprises the SPM Sunphotometers and also two types of absolute radiometers (VIRGO/DIARAD 
and VIRGO/PMO6-V) and one Luminosity Oscillation Imager (VIRGO/LOI). In this section we study the 220.7 $\mu$Hz signal in these instruments.

\begin {itemize}
\item{Luminosity Oscillation Imager (VIRGO/LOI)}

VIRGO/LOI measures  the radiance in 12 pixels over the solar disc. We convert these 12 pixels into one by simply adding all of them. From 
the raw time series the same analysis as in VIRGO/SPM has been carried out.  Figure~\ref{loi} ({\it Top}) shows its time-evolution diagram. 
It looks similar to the VIRGO/SPM and with the same visible signal at 220.7 $\mu$Hz. In the VIRGO/LOI observations, the signal is weaker 
than in VIRGO/SPM but with the same characteristics, for example, at time series 60 the signal slightly changes its frequency. However, 
with this instrument, the peak seems to be like a doublet instead of only one signal concentrated in a couple of bins.
\begin{figure}[]
\centering
\begin{tabular}{c}
	\includegraphics[width=12pc,angle =90]{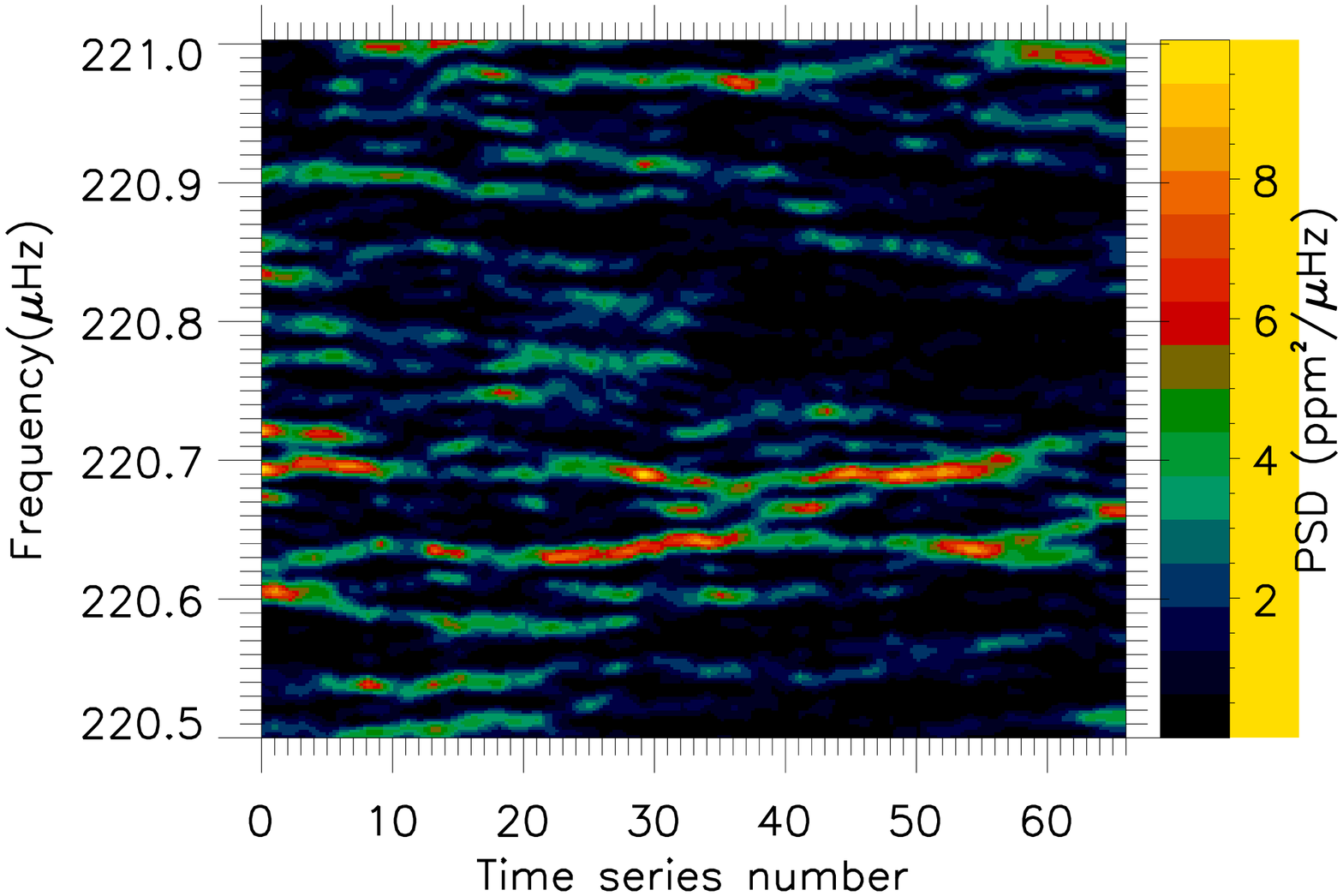} \\
  	\includegraphics[width=12pc,angle =90]{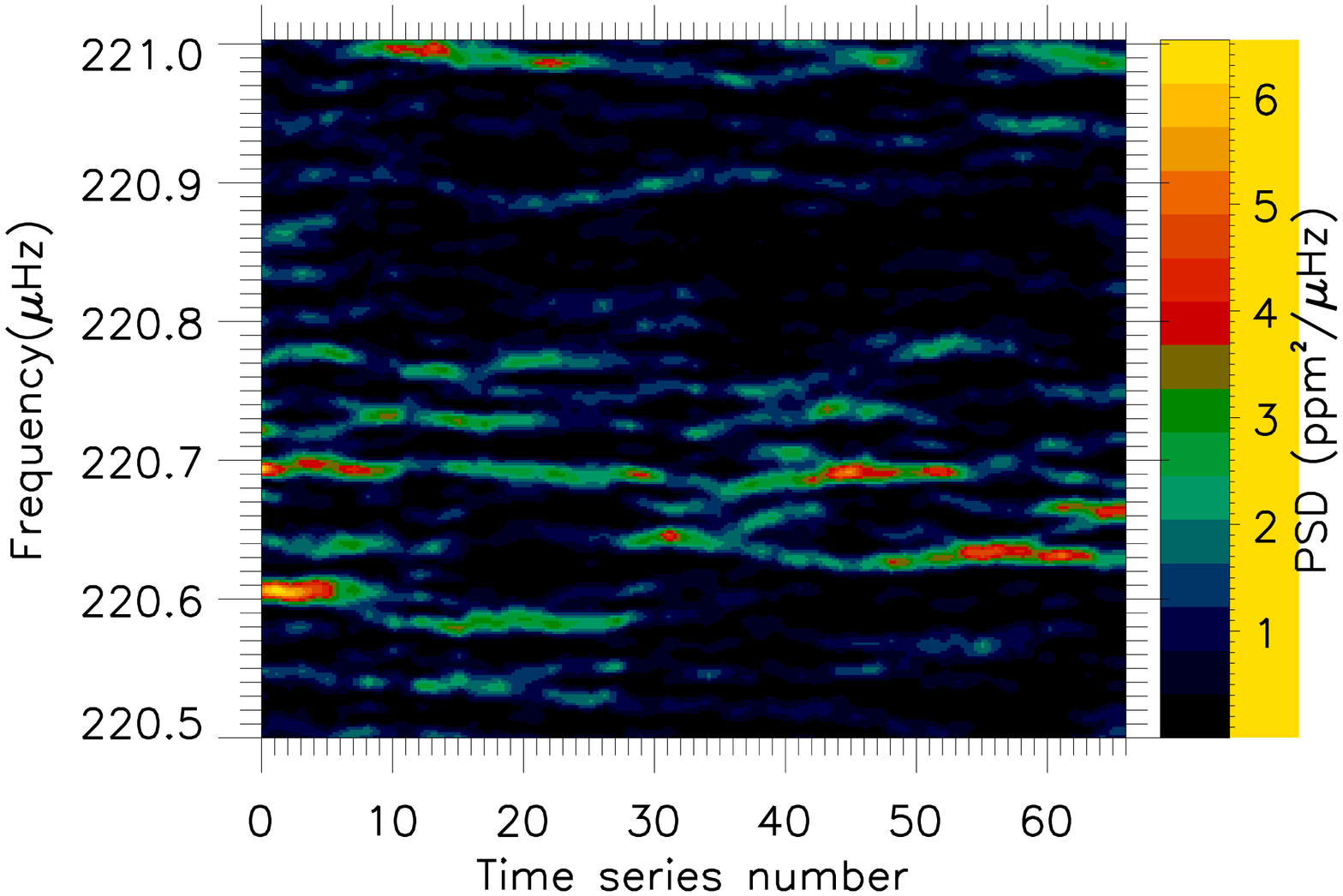} \\
	\includegraphics[width=12pc,angle =90]{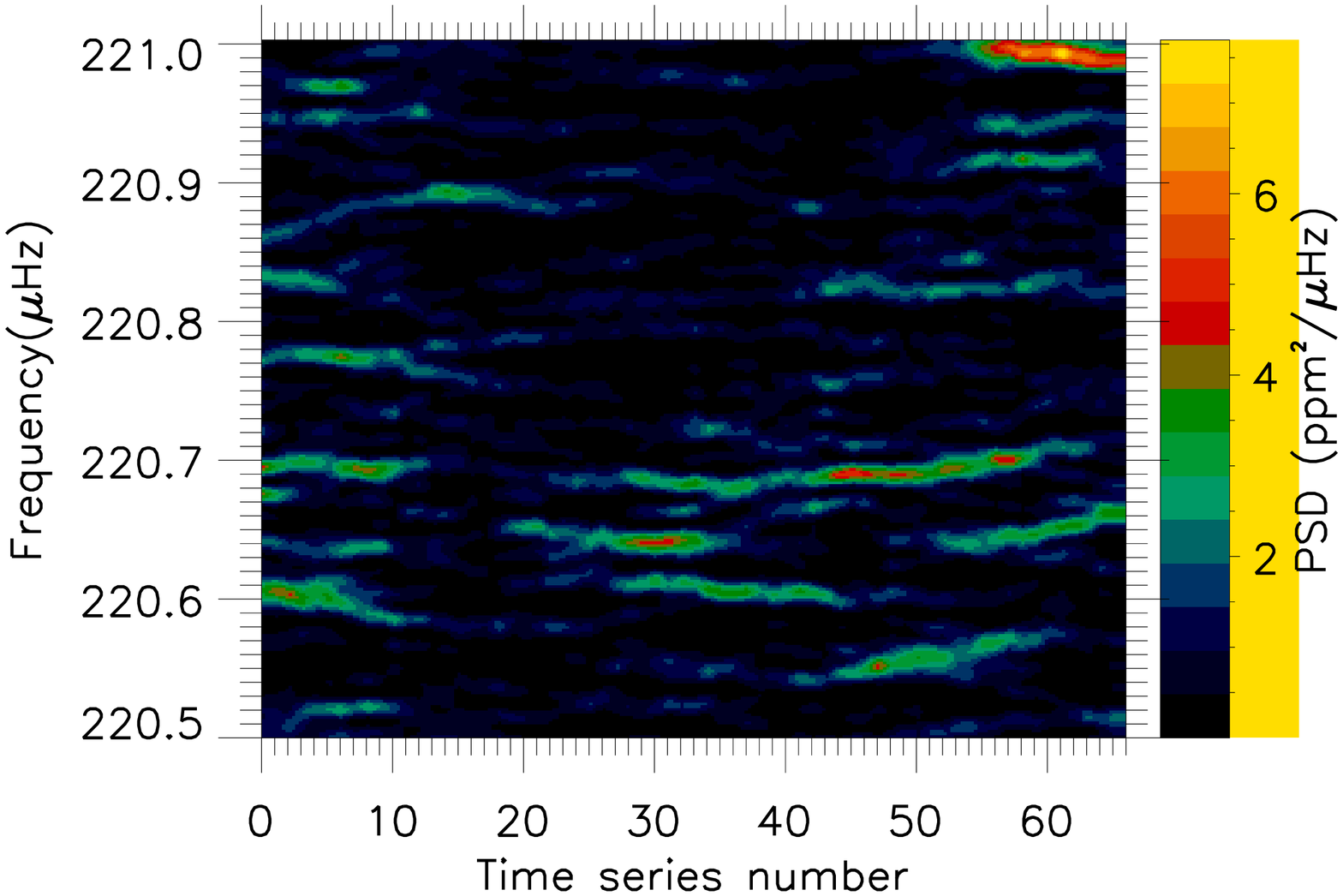} \\
\end{tabular}
\caption{\label{loi}Time-evolution power diagram of the Luminosity Oscillation Imager (VIRGO/LOI),  the VIRGO/DIARAD absolute radiometer 
and the VIRGO/PMO6-V absolute radiometer, respectively from top to bottom. All these instruments are part of the VIRGO package.}
\end{figure} 

\item{Absolute radiometers (VIRGO/DIARAD and VIRGO/PMO6-V)}

Absolute radiometers use a quite different technique from that of VIRGO/SPM and VIRGO/LOI, which are silicon detectors measuring the 
spectral irradiance and the radiance respectively. Absolute radiometers are based on the measurements of the heat flux by using an 
electrically calibrated heat flux transducer to measure the total solar irradiance (solar constant). Once again, from the raw time 
series we have performed the same analysis. Figure~\ref{loi} ({\it medium and bottom}) shows the results for VIRGO/DIARAD and VIRGO/PMO6-V. 
Time-evolution power diagrams are similar to the previous ones, although the 220.7 $\mu$Hz is weaker in both radiometers but the 220.7 $\mu$Hz signal is still present.

\end{itemize}

\section{Analysis using Doppler velocity instrumentation}

Up to now we have not found any instrumental origin for the 220.7 $\mu$Hz signal observed in all the VIRGO package. We can now study this 
region in other helioseismic instruments. We will start by analysing the signal of the other two instruments on board SoHO and we finish by
 using the GONG ground-based network.

\begin{itemize}
\item{GOLF} is the other Sun-as-a-star instrument on board SoHO. We have analysed the velocity time series following the same procedure 
employed in the VIRGO analysis and we have computed the time-evolution power diagram shown in Figure \ref{Doppler} {\it (top)}. As  
 mentioned in the introduction, the 220.7 $\mu$Hz signal was first observed by GOLF during the first years of the mission and it was flagged 
as a ``g-mode'' candidate by \cite{STC04} and, after 4182 days, it is still visible as part of a quadruplet above a 90\% confidence level 
\citep{2008AN....329..476G}. Figure~\ref{Doppler} {\it (Top)} shows that the evolution with time of the signal, although weaker than in VIRGO, 
is still there. The signal in GOLF has an interval between time series 12 and 19, where it disappears, and it corresponds to the place  where 
the signal in VIRGO/SPM is the weakest (see figure~\ref{spms}). Therefore, we can conclude that the 220.7 $\mu$Hz is also observed in velocity 
measurements using GOLF data but with a smaller signal-to-noise ratio (SNR).
\begin{figure}[]
\centering
\begin{tabular}{c}
	\includegraphics[width=12pc,angle =90]{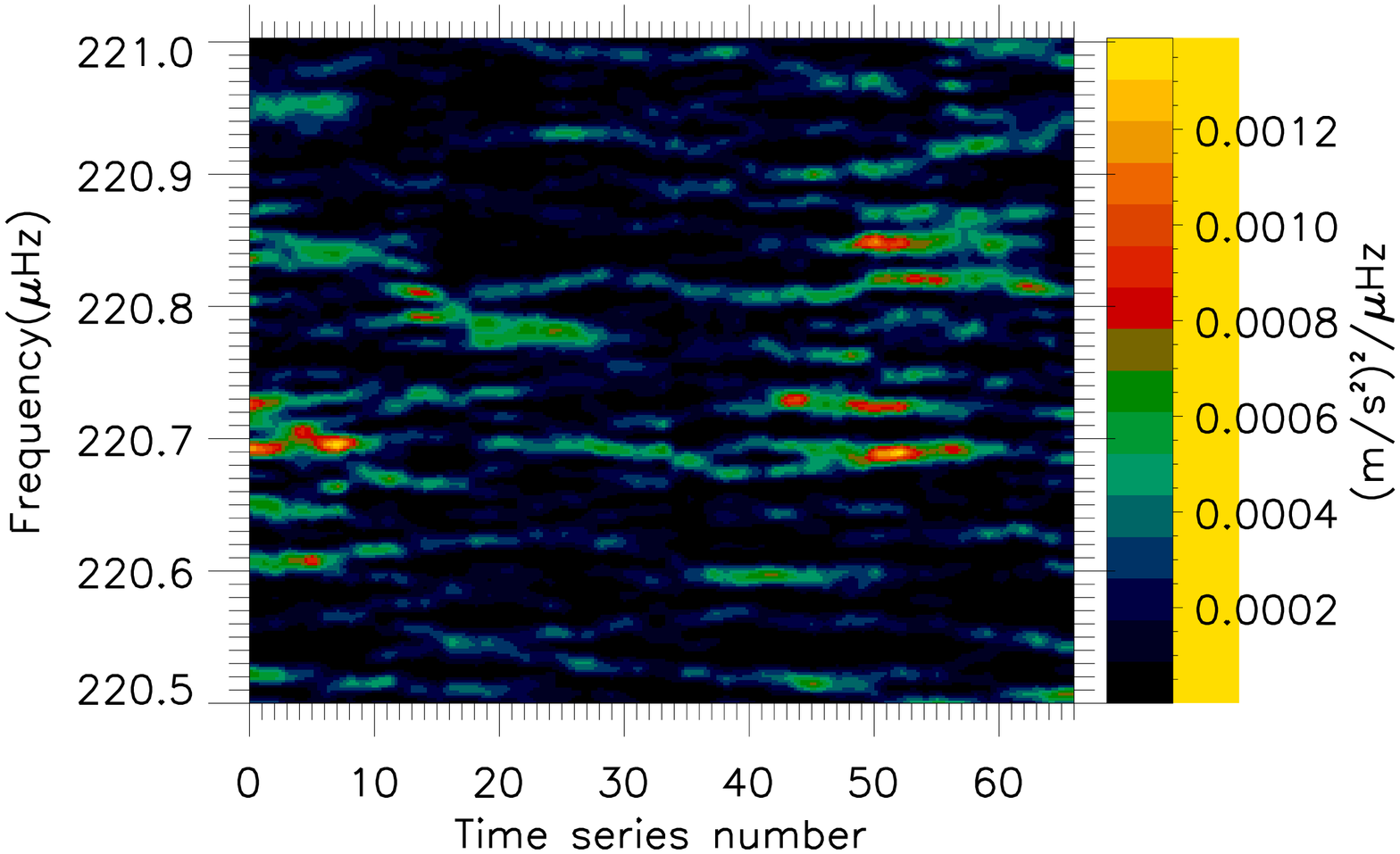} \\
  	\includegraphics[width=12pc,angle =90]{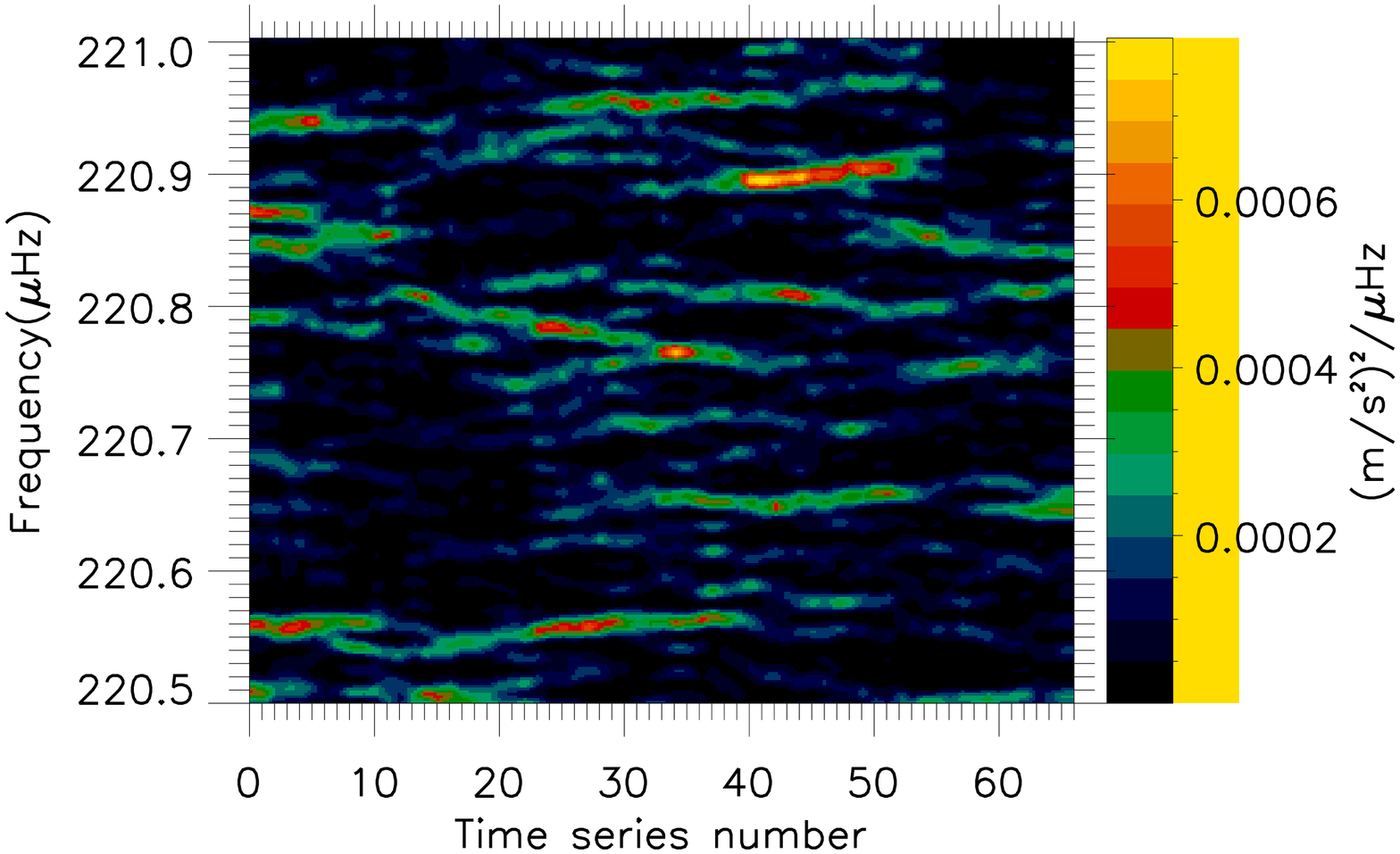} \\
	\includegraphics[width=12pc,angle =90]{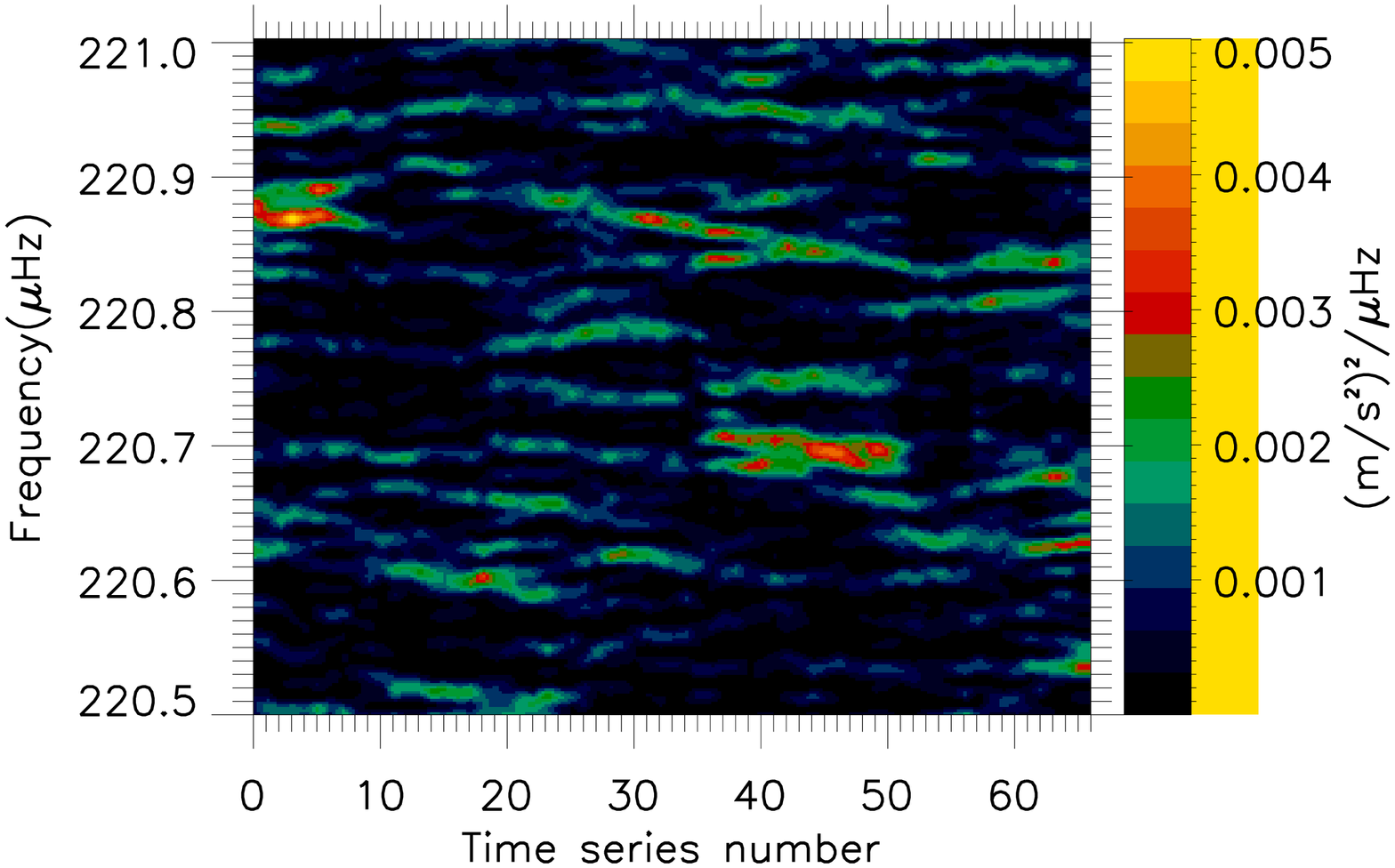} \\
\end{tabular}
\caption{\label{Doppler}Time-evolution power diagram of the GOLF, MDI and GONG instruments respectively from top to bottom.}
\end{figure} 

\item{}Disc-averaged MDI velocity signals from the calibrated level-1.4 MDI LOI-proxy Doppler images were obtained using integrated spatially 
weighted masks following \cite{Henney99}. These time series from 1996 May 25 till  2007 October 28 have been analysed and the time-evolution 
diagram plotted in Figure~\ref{Doppler} {\it (medium)}. There are no fingerprints of the presence of the 220.7 $\mu$Hz signal in this data set. 
This could be due to a lower SNR in the MDI LOI-proxy as compared to GOLF. Indeed, \cite{HenneyUlrich99} showed that for the lowest measurable 
p-modes, the GOLF instrument has a higher SNR than this particular MDI mask. To go further, we need to apply our methodology to specifically 
designed g-mode masks such as those derived by \cite{Watcher}.

\item{}The radial velocity of GONG used in this work started on 1995 May 7 and finished on 2006 March 9. These series are shifted by a year compared to the 
SoHO data but they were the longest we could use. The disc-integrated data provided by the GONG Team are very noisy at low frequency and were not 
suited for our studies. Therefore, we have decided to use the $\ell$=2 spherical-harmonic series. These are optimized for acoustic modes of this 
degree and they have the advantage of having a much stabler behaviour at low frequency. We preferred these series to the $\ell$=1 mode because the 
closest theoretical frequency to the target frequency of 220.7 $\mu$Hz corresponds to an $\ell$=2 g mode. In Figure~\ref{Doppler} {\it(bottom)} 
the time-evolution diagram of GONG is shown. Although the SNR at this frequencies in the GONG data is very low, it seems to be a trace of the 
220.7 $\mu$Hz signal in the GONG data, especially between series 36 and 52, which corresponds to a maximum in the GOLF between series 43 and 60 
(7 subseries shift, i.e.\ $\sim$ 350 days). However, looking only at this time-evolution diagram, it is impossible to disentangle the feature at 
220.7 $\mu$Hz from others visible in the analysed region. Nevertheless, It would be extremely important to be able to confirm the visibility of 
such a peak using ground-based data because that would directly mean that the nature of this 220.7 $\mu$Hz signal has a solar origin.
\end{itemize}

\begin{figure}[!hbt]
\centerline{%
\begin{tabular}{c@{\hspace{1pc}}c}
\includegraphics[width=18pc,angle =0]{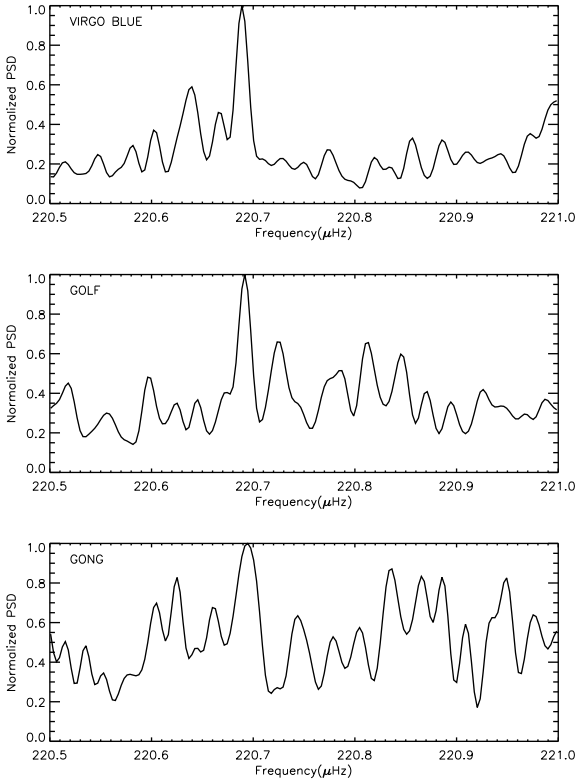}
\end{tabular}}
\caption{\label{Colapso} Collapsograms of the time-evolution power diagrams of VIRGO/blue, GOLF and GONG. The 220.7  $\mu$Hz structure is
 present in the three different instruments although in GONG it is at noise level.}
\end{figure}

To compare the averaged behavior of the 220.7 $\mu$Hz signal in the GOLF, GONG and VIRGO/SPM data sets we  computed the collapsograms of the 
time-evolution power diagram, i.e.\ to average the 66 power spectra used to produce the time-evolution power diagrams. The resultant graphs 
are plotted in Figure~\ref{Colapso}. A similar structure appears around  the target frequency of  220.7 $\mu$Hz,   this peak being the highest in 
the three instruments. However, in the case of GONG, it is at noise level. 

\section{Conclusions} 
In the present paper we have studied a peak that appears around the frequency of 220.7 $\mu$Hz in the VIRGO/SPM data. This peak has a more 
than 90$\%$ confidence level  of not being due to noise in the full spectrum of 4098 days. A detailed study of its nature revealed that this 
peak existed since the very begining of the mission in a continuous way for the last 11 years and only at the very end of the  
time series considered does it seem to  change slightly in frequency. By Monte Carlo simulations we have computed the confidence level of such kinds of behaviour
 and we found that it is really unlikely (more than 99$\%$) that it is due to a noise with the same statistical characteristics as the convective
  noise. Therefore, we  checked all the available housekeeping data from the VIRGO package as well as a detailed analysis of the SoHO 
  spacecraft attitude control, looking for an instrumental origin. None of these studies was able to explain the presence of a peak in the  
  region studied. Indeed, this study seems to rule out this possibility. The  origin should therefore be solar. We  then studied Doppler velocity data 
  from another instrument on board SoHO, GOLF, and we  found that the peak is also present (with lower SNR). Even though  analysis of data from the GONG 
  ground-based network revealed a very noisy spectrum, the highest peak in a 10 $
\mu$Hz region around the 220.7 $\mu$Hz signal is precisely that peak. However, it is not significant enough for us to claim that we have a positive 
detection using this instrument.

The present study has proved the solar origin of the peak at 220.7 $\mu$Hz. Two solar phenomena could be responsible of such a peak. The first could be 
 convection, in particular  granulation motions. However, it is very unlikely that a turbulent displacement of plasma on the solar surface 
with a typical time scale of 10 minutes give a stable frequency during more than 10 years in the power spectrum of the disc-integrated data. On 
the other hand, gravity modes propagate inside the radiative region of the Sun and  are expected to have long lifetimes (at least longer 
than the period of measurements). Thus, the properties of the peak that we found are similar to those expected for a g mode. Using the principle of Ockham's 
razor  (or the {\it Lex Parsimoniae} principle) in which the explanation of any phenomenon should make as few assumptions as possible, we 
can conclude that if this peak is not noise it should be a component of a g mode.  Analysing in detail the structure of this possible g-mode 
component, Figure~2 reveals a peak structure containing several bins. Indeed, Figure~1 might also show the presence of a parallel component at 
around 220.64 $\mu$Hz with high amplitudes in several of the  series considered. Thus, a possible explanation of such behaviour might be the presence 
of an inner magnetic field that could slightly split the component of the g-mode multiplet in some peaks. Another possibility might be that the 
g-mode power could be spread into several bins as a consequence of a smaller  than expected lifetime or due to a change in the size of the resonant 
cavity (for example due to a displacement of the position of the tachocline during the activity cycle). This latter effect is particularly interesting 
because it seems that the 220.7 signal follows a small change in frequency over the entire time-span with the lowest frequency (220.68 $\mu$Hz) 
reached around time series 35---corresponding to the maximum of the activity cycle--- and then increasing the frequency again towards the two 
periods with minimum activity (at the beginning and the end of the series). In any case,  assuming a faster rotation in the core than in the 
rest of the radiative envelope (as suggested by Garc\'\i a et al. 2007), the 220.7 $\mu$Hz peak could be or a component of the $\ell$=2, n=-3 g
 mode, or a component of the $\ell$=3, n=-5 or a bitting between this latter and the $\ell$=5, n=-8. Whatever the true answer is, there is still 
 an important question to be answered: why is this particular peak so excited when there are no other visible g-mode components? More work will be 
 necessary before solving the solar g-mode puzzle.

\begin{acknowledgements}
The authors want to thank the members of the PHOEBUS group present at the first ISSI (International Space Science Institute) meetings for 
their useful comments and discussions. This work has been partially supported by the Spanish grant PNAyA2007-62651 and the CNES/GOLF grant at the SAp/CEA-Saclay.
The authors also thank all their colleagues (scientists, engineers and technicians) involved with the GOLF, VIRGO and MDI instruments
 aboard SoHO which is a space mission of international cooperation between ESA and NASA. This work utilizes data obtained by the Global 
 Oscillation Network Group (GONG) program, managed by the National Solar Observatory, which is operated by AURA, Inc. under a cooperative 
 agreement with the National Science Foundation. The data were acquired by instruments operated by the Big Bear Solar Observatory, High Altitude 
 Observatory, Learmonth Solar Observatory, Udaipur Solar Observatory, Instituto de Astrof\'\i sica de Canarias, and Cerro Tololo Interamerican Observatory. One of the authors (AJ) would like to thank M.Ortiz for invaluable help with the last part of this article. 
 
\end{acknowledgements}

\end{document}